\documentclass[a4paper,12pt]{article}
\usepackage{amsmath,amsthm,amssymb,amscd}
\usepackage[margin=1in]{geometry}
  \def\NN{\mathbf N}
\def\x{\mathbf x} \def\bz{\mathbf z}
\let\d\partial
\let\a\alpha \let\ph\varphi  \let\o\omega 
\let\de\delta \let\l\lambda \let\a\alpha \let\be\beta \let\ga\gamma
 \let\t\tau \let\ep\epsilon
\let\p\pi \def\tp{\tilde\p}
\def\Hb{\,\overline{\!H}}
\def\Vb{\overline V}
\def\Qt{\tilde Q}
\def\ct{\tilde c}
\def\g{\mathfrak g}
\def\sL{\mathfrak{sl}}
\def\borel{\mathfrak b_{N+1}}
\def\e{\mathrm e} \def\ii{\mathrm i}
\def\E{\mathcal E}
\def\U{\mathcal U}
\def\N{\mathcal N}
\def\L{\mathcal L}
\def\M{\mathcal M}
\def\Mb{\,\overline{\!\mathcal M}}
\def\D{\mathcal D}
\let\ds\displaystyle
\def\abs#1{\left|#1\right|}
\def\operator#1{\expandafter\def\csname#1\endcsname{\operatorname{#1}}}
\operator{sech} \operator{csch}

\newtheorem{thm}{Theorem}
\newtheorem{lemma}[thm]{Lemma}
\theoremstyle{definition}

\theoremstyle{remark}

\def\({{\rm(}} \def\){{\rm)}}
\begin{document}
\title{New Algebraic Quantum Many-body Problems%
\thanks{Supported in part by DGES Grant PB98--0821.}}
\author{\large D.
G\'omez-Ullate\footnote{dgu@eucmos.sim.ucm.es}\,,\enspace A.
Gonz\'alez-L\'opez\footnote{artemio@eucmos.sim.ucm.es}\enspace
and\enspace M. A. Rodr\'{\i}guez\footnote{rodrigue@eucmos.sim.ucm.es}
\\[12pt] \normalsize
\begin{tabular}{c}
Departamento de F\'{\i}sica Te\'orica II\\
Facultad de Ciencias F\'{\i}sicas\\
Universidad Complutense\\
28040 Madrid, Spain
\end{tabular}}
\date{March 1, 2000}
\maketitle
\begin{abstract}
We develop a systematic procedure for constructing quantum many-body
problems whose spectrum can be partially or totally computed by
purely algebraic means. The exactly-solvable models include
rational and hyperbolic potentials related to root systems, in
some cases with an additional external field. The quasi-exactly
solvable models can be considered as deformations of the previous
ones which share their algebraic character.
\end{abstract}
\vfill
Submitted to \emph{J. Phys. A}
\vspace{1cm}
\newpage
%
%
\section{Introduction}
\label{intro}
Since the pioneering works of Calogero and Sutherland,
\cite{Cal1,Cal2,Suth}, much effort has been invested in the study
of quantum many-body problems. A thorough classification of
completely integrable Hamiltonians related to root systems
has been performed in the early eighties by Olshanetsky and
Perelomov, both in the classical, \cite{OPclass}, and the
quantum cases, \cite{OP}.
The complete integrability of these models is associated to an
underlying root system structure, the integrals of motion being
related to the radial parts of the Laplace--Beltrami operator on a
symmetric space with the given root system.

Some years ago, Turbiner showed using Lie-algebraic techniques that
many (though not all) of Olshanetsky--Perelomov's (OP) Hamiltonians
are also exactly solvable, \cite{Turb1,Turb3}. By exact solvability we
mean here that the $N$-body Hamiltonian preserves an infinite
increasing sequence of subspaces of known smooth functions, whereas a
quasi-exactly solvable Hamiltonian just preserves a single
finite-dimensional subspace. The construction of these invariant
subspaces is usually based on the theory of representations of Lie
algebras of differential operators, \cite{GKOJPA}.

The idea of constructing exactly solvable many-body problems based
on the zeros and poles of special solutions of partial
differential equations has long been known. The original work
\cite{Caltruco} explores in depth this relation at the classical
level. This idea has been exploited by Hou and Shifman in a recent
paper, \cite{Shifman}, which extends this approach to
the quantum case using Lie-algebraic techniques, thereby obtaining
a new family of quasi-exactly solvable many-body potentials.

In this paper we apply Calogero's procedure for constructing solvable
many-body problems to the most general quasi-exactly solvable operator
on the line admitting square-integrable eigenfunctions, finding
several families of exactly and quasi-exactly solvable many-body
problems on the line. The exactly solvable Hamiltonians include
Calogero's original model and many of Olshanetsky--Perelomov's
non-periodic Hamiltonians, and also a family of potentials which are
not directly related to a root system due to the presence of an
external field. The latter model has been studied by Inozemtsev and
Meshcheryakov, \cite{Inocencio}, who derived its discrete spectrum 
from that of the standard hyperbolic $BC_N$ model by a limiting 
process.

It should be emphasized that, although the construction presented here
leads to many previously known potentials based on root systems, our
approach is entirely different from the usual one, in that no use
needs to be made of any underlying root system structure. The main
advantage of this algebraic method over the root system approach is
that, while the latter is rather rigid, the former's flexibility
allows for deformations of the exactly solvable models preserving to
some extent their integrability properties.

The paper is organized as follows. In Section \ref{der} some basic
definitions are given, and the construction of the general many-body
Hamiltonian is explained. The algebraization of this Hamiltonian is
discussed in Section \ref{alghn} for the five families of normalizable
quasi-exactly solvable operators on the line. In Section \ref{EnSp}
the energy spectrum of all exactly-solvable models obtained in the
previous Section eis given explicitly. In Section \ref{exa} we work out
a few concrete examples, and we sum up the conclusions and outline
future work in Section \ref{sumcon}. Some useful expressions for
handling symmetric variables can be found in the Appendix.
%
%
\section{Derivation of $H_N$}
\label{der}
A Schr\"odinger operator
\begin{equation}
    H=-\sum_k \d_{x_k}^2+V(\x)
    \label{schr}
\end{equation}
(where $\x$ belongs to an open subset of Euclidean space) is said to
be \emph{quasi-exactly solvable} (QES) if it preserves a known
finite-dimensional subspace $\M$ of smooth functions. The spectral
problem for $H$ reduces in this space to diagonalizing the matrix of
$H|_{\M}$, which makes it possible to compute a finite subset of the
spectrum of $H$ by purely algebraic means. If $H$ preserves an
infinite increasing sequence of known finite-dimensional subspaces
$\M_0\subset\M_1\subset\dots\subset\M_k\subset\dots$ then we shall
call it \emph{exactly solvable} (ES), since in this case one can
algebraically compute an arbitrary number of eigenfunctions and
eigenvalues of $H$ by restricting it to each $\M_k$. (If $\M$ or the
$\M_k$'s are not in $L^2$, some of the eigenfunctions of $H$ computed
in this way may turn out not to be square-integrable; see
\cite{Artemio93} for an analysis of this issue in the one-dimensional
case.)

The QES (or ES) character of a Schr\"odinger operator is invariant
under a natural group of equivalence transformations, generated by
changes of coordinates $\x\mapsto \bz=\mathbf Z (\x)$ and
conjugation by arbitrary non-negative
functions $\mu(\bz)$, which map $H\equiv H(\x)$ into the differential
operator (not necessarily of Schr\"odinger type)
\begin{equation}
    \Hb(\bz)=\mu(\bz)^{-1} H(\x)\,\mu(\bz)\,.
    \label{gh}
\end{equation}
Thus, if $H(\x)$ preserves $\M\equiv\M(\x)$ then $\Hb(\bz)$ preserves
the subspace
\begin{equation}
    \Mb(\bz)=\mu(\bz)^{-1} \M(\x)\,,
    \label{mub}
\end{equation}
and if $\psi(\x)$ is an
eigenfunction of $H(\x)$ with eigenvalue $E$ belonging to $\M(\x)$ then
$\overline\psi(\bz)=\mu(\bz)^{-1} \psi(\x)$ is an eigenfunction of
$\Hb(\bz)$ lying in $\Mb(\bz)$ with the same energy $E$. Note, however,
that such an equivalence transformation may not preserve the square
integrability of the eigenfunctions, since $\mu(\bz)$ is not required
to be unimodular.

A very general way of constructing QES Schr\"odinger operators is to
start with a finite-dimensional Lie algebra of differential operators
$\bar\g$ admitting finite-dimensional representations in the space of
smooth functions in the variable $\bz$ (usually called a
\emph{quasi-exactly solvable algebra} in the literature). Any
differential operator $\Hb(\bz)$ belonging to the enveloping algebra
of $\bar\g$ will automatically preserve the carrier space $\Mb(\bz)$
of such a finite-dimensional representation. If one can find an
equivalence transformation \eqref{gh} such that the differential
operator $H(\x)$ is of Schr\"odinger type \eqref{schr}, then $H(\x)$
is clearly a QES Schr\"odinger operator, since it preserves the
finite-dimensional subspace $\M(\x)=\mu(\bz)\,\Mb(\bz)$. In this case,
one says that $\g=\mu(\bz)\cdot\bar\g\cdot\mu(\bz)^{-1}$ is the
\emph{hidden symmetry algebra} responsible for the QES character of
$H(\x)$. In this paper we shall be almost exclusively concerned with
this special type of quasi-exactly solvable Schr\"odinger operators,
that we shall call \emph{algebraic} to single them out from the rest.
The function $\mu(\bz)$ is usually called the \emph{gauge factor} in
the literature, and $\Hb(\bz)$ is referred to as the \emph{gauge
Hamiltonian.}

In one dimension, the only Lie algebra of first-order
differential operators is (up to equivalence) the standard projective
realization of $\sL(2)$ (or its
subalgebras), with basis elements
\begin{equation}\label{sl2}
J_N^- = \d_z, \qquad J_N^0 = z\, \d_z - \frac{N}{2},\qquad J_N^+ =
z^2\,
\d_z - N\,z\,.
\end{equation}
If $N$ is a non-negative integer, the latter algebra admits
a $(N+1)$-dimensional representation in the space $\mathcal P_N$ of
polynomials of degree $\leq N$. The most general second-order
differential operator belonging to the enveloping algebra of the Lie
algebra \eqref{sl2},
obtained by constructing an arbitrary quadratic combination of the
generators (\ref{sl2}), is of the form
\begin{equation}\label{QESgen2}
-\Hb(z) = P(z)\,\d_z^2 + \tilde Q(z)\,\d_z + \tilde R(z)
\end{equation}
with
\begin{equation}
    \Qt(z)=Q(z)-\frac{N-1}2\,P'(z)\,,\qquad \tilde R(z) = R-\frac N2\,
    Q'(z)+\frac
    N{12}(N-1)P''(z)\,,
    \label{qtrt}
\end{equation}
where $P$, $Q$ and $R$ are arbitrary polynomials of degree $4$, $2$
and $0$, respectively, and the minus sign is for later convenience. If
$P$ is positive, the operator \eqref{QESgen2} is equivalent (in the
sense of \eqref{gh}) to a Schr\"odinger operator $H(x)=-\d_x^2+V(x)$
by the change of variables \begin{equation}\label{cofv} x=\xi(z) =
\int^z \frac{dy}{\sqrt{P(y)}}
\end{equation}
and conjugation by the gauge factor
\begin{equation}\label{gauge1}
\mu(z) = P(z) ^{-1/4} \exp \left\{ \int^z \frac{ \tilde
Q(y)}{2P(y)}\,dy \right \}.
\end{equation}
Let $N\ge1$, and assume, for the sake of simplicity, that the gauge
Hamiltonian \eqref{QESgen2} is diagonalizable in $\mathcal P_N$. Then
$\Hb(z)$ has $N+1$ algebraically computable \emph{polynomial}
eigenfunctions $\ph_k(z)$ ($0\le k\le N$) of degree $\leq N$, and
therefore $H(x)$ has $N+1$ algebraically computable eigenfunctions of the
form
\begin{equation}
    \psi_k(x)=\mu(z)\,\ph_k(z)|_{z=\xi^{-1}(x)}\,,
    \qquad 0\le k\le N\,.
    \label{algeig}
\end{equation}

Following Calogero's original idea, consider now the
time-dependent Schr\"odinger equation with Hamiltonian $H(x)$, namely
\begin{equation}\label{TDSE}
H (x) \Psi(x,t) = \ii\,\d_t \Psi(x,t)\,.
\end{equation}
Since $H(x)$ is time-independent, the latter equation
will admit solutions of the form
\begin{equation}
    \label{Psix}
\Psi(x,t) = \sum_{k=0}^N c_k\,\psi_k(x)\,\e^{-\ii E_k t}\,,
\end{equation}
where $\psi_k(x)$ is an eigenfunction of $H(x)$ of the form
\eqref{algeig} with
energy $E_k$, and $c_0,\dots,c_N$ are arbitrary complex constants.
Equivalently,
\begin{equation}
    \Psi(x,t)=
    \mu(z)\Phi(z,t)|_{z=\xi^{-1}(x)}\,,
    \label{Psiz}
\end{equation}
with
\begin{equation}
    \Phi(z,t)=\sum_{k=0}^N c_k\,\ph_k(z)\,\e^{-\ii E_k t}\,.
    \label{Phi}
\end{equation}
Since each $\ph_k(z)$ is a polynomial of degree $\leq N$ (and at least
one of them is of degree $N$, since otherwise $\Hb(z)$ would not be
diagonalizable in $\mathcal P_N$), it follows from the latter equation
that $\Phi(z,t)$ is a polynomial of degree $N$ in $z$ with
time-dependent coefficients. Therefore we can write
\begin{equation}\label{varphi}
\Phi(z,t) = C(t)\,\prod_{j=1}^N [z-z_j(t)]\,.
\end{equation}
It follows from Eq.~\eqref{Psiz} that each zero $z_j(t)$ of
$\Phi(z,t)$ yields a zero
\begin{equation}
    x_j(t)=\xi\bigl(z_j(t)\bigr)\,,\qquad 1\le j\le N\,,
    \label{xjt}
\end{equation}
of the time-dependent wavefunction $\Psi(x,t)$. We will study the
motion of these zeros, showing that it can be derived from a
classical Lagrangian.

Indeed, substituting \eqref{Psiz} into Schr\"odinger's
equation \eqref{TDSE} and using \eqref{varphi} we easily obtain the
equation
\begin{equation}\label{eqzt}
\sum_{\substack{j,k=1\\j \ne k}}^N
\frac{P(z)}{\bigl(z-z_j(t)\bigr)\bigl(z-z_k(t)\bigr)}
+ \sum_{j=1}^N
\frac{\Qt(z)}{z-z_j(t)} + \tilde R(z) =
\ii\,\sum_{j=1}^N \frac{\dot z_j(t)}{z-z_j(t)}-
\ii\,\frac{\dot C(t)}{C(t)}\,,
\end{equation}
which must hold identically in $z$ and $t$. Equating the residue of
both sides at $z=z_k(t)$ we arrive at the following system of
differential equations for the functions $z_k(t)$:
\begin{equation}\label{residue}
\ii\,\dot z_k = \Qt(z_k) + 2 \sum_{\substack{j=1\\j\ne k}}^N
\frac{P(z_k)}{z_k-z_j} \equiv F_k(\bz) \,,\quad 1\leq k\leq N\,.
\end{equation}
Conversely, it can be shown without difficulty that the latter
equations imply Eq.~\eqref{eqzt}. In order to show that
$(x_1(t),\dots,x_N(t))$ follows a trajectory of a certain
Lagrangian system, we shall make use of the following Lemma:
\begin{lemma}
Consider the autonomous system of first-order ordinary differential
equations
\begin{equation} \label{dotx}
\ii\,\dot x_k = f_k(\x);\qquad k=1,\dots,N\,,\quad
\x\equiv(x_1,\dots,x_N)\in\bf C^N\,.
\end{equation}
If the one-form $\sum_{k=1}^N f_k(\x)\,dx_k$ is closed, then the
second-order system obtained by differentiating \eqref{dotx}
once with respect to $t$ is Lagrangian, with Lagrangian given by
\begin{equation}
    \label{lag}
L = \dot \x^2-\sum_{j=1}^N f_j^2(\x)\,.
\end{equation}
\end{lemma}
\begin{proof}
The second-order system obtained by differentiating
\eqref{dotx} with respect to $t$ is
\begin{equation}\label{second}
\ddot x_k=-\ii\,\sum_{k=1}^N \dot x_j\,\frac{\d f_k}{\d x_j}
=-\sum_{k=1}^N f_j\,\frac{\d f_k}{\d x_j}\,,\qquad 1\le k\le N\,.
\end{equation}
On the other hand, the Euler--Lagrange equations associated to the
Lagrangian \eqref{lag} are
\begin{equation}\label{EL}
2\,\ddot x_k=\frac{\d L}{\d x_k}=-2\,\sum_{j=1}^N f_j\,\frac{\d
f_j}{\d x_k}\,, \qquad 1\le k\le N\,.
\end{equation}
If $\sum_{k=1}^N f_k(\x)\,dx_k$ is closed then
$$
\frac{\d f_k}{\d
x_j}=\frac{\d f_j}{\d x_k}\,,
$$
so that \eqref{second} and
\eqref{EL} are indeed identical.
\end{proof}
In our case, the system (\ref{dotx}) is obtained from
(\ref{residue}) by the change of independent variables \eqref{xjt},
with $\xi(z)$ as in (\ref{cofv}). Since $\dot x_k = {P(z_k)}^{-1/2}\,\dot
z_k$, we have
\begin{equation}
    f_k={P(z_k)}^{-1/2}\,F_k
    \label{fk}
\end{equation}
and
\begin{equation}
\sum_{k=1}^N f_k(\x)\,dx_k = \sum_{k=1}^N
\frac{F_k(\bz)}{P(z_k)}\,dz_k\,.
\end{equation}
The latter one-form is clearly closed, since if $j\ne k$ we have
$$
\frac{\d}{\d
z_j}\left(\frac{F_k(\bz)}{P(z_k)}\right)= \frac{\d}{\d
z_j}\left[\frac{\tilde Q(z_k)}{P(z_k)}+
\sum_{\substack{l=1\\l\ne k}}^N\frac2{z_k-z_l}\right] =
\frac2{(z_k-z_j)^2}\,,
$$
which is clearly symmetric under the exchange of $j$ and $k$. It
follows from the previous Lemma and Eqs.~\eqref{lag} and \eqref{fk}
that the zeros $\x(t)$ of the time-dependent wavefunction \eqref{Psix}
move along a trajectory of the Lagrangian system with Lagrangian
\begin{equation}
L = \sum_{k=1}^N \dot x_k^2 - \sum_{k=1}^N
\frac{F_k^2(\bz)}{P(z_k)}\,,
\end{equation}
whose associated Hamiltonian is
\begin{equation}\label{ham1}
H_N = \sum_{k=1}^N p_k^2 + \sum_{k=1}^N \frac{F_k^2(\bz)}{P(z_k)}\,.
\end{equation}
Substituting (\ref{residue}) into (\ref{ham1}) and letting
$p_k=-\ii\,\d_{x_k}$ we obtain the $N$-particle
quantum Hamiltonian
\begin{equation}\label{ham2}
H_N =-\sum_{k=1}^N \d_{x_k}^2+ V_N (\x)\,,
\end{equation}
where
\begin{equation}
    \label{VN0}
 V_N (\x)= \sum_{k=1}^N \frac{\Qt_k^2}{P_k} + 4
 \sum_{\substack{j,k=1\\j\ne
 k}}^N
\frac{\Qt_k}{z_{kj}} + 4 \sum_{\substack{j,k=1\\j\ne
 k}}^N
\frac{P_k}{z_{kj}^2} + 4\!\!\sum_{\substack{j,k,l=1\\j\ne
 k\ne l\ne j}}^N
\frac{P_k}{z_{kj}z_{kl}}
\end{equation}
and we have set
$$
P_k= P(z_k)\,,\qquad\Qt_k = \Qt(z_k)\,,
\qquad z_{kj}= z_k-z_j\,.
$$
It will be convenient for our purposes to allow the coefficients of
each of the four sums in Eq.~\eqref{VN0} to be different. Dropping an
inessential overall factor we finally arrive at the following formula
for the potential:
\begin{equation}\label{Vn}
V_N(\x) =  \frac{1}{4}\sum_{k=1}^N \frac{\tilde Q_k^2}{P_k} + g_1
\sum_{\substack{j,k=1\\j\ne
 k}}^N\frac{\tilde Q_k}{z_{kj}} + g_2 \sum_{\substack{j,k=1\\j\ne
 k}}^N
\frac{P_k}{z_{kj}^2}+g_3\!\!\sum_{\substack{j,k,l=1\\j\ne
 k\ne l\ne j}}^N
\frac{P_k}{z_{kj}z_{kl}}\,.
\end{equation}

The main purpose of this paper is to show that, for certain choices of
the polynomials $P(z)$ and $\Qt(z)$ (and, sometimes, the constant
$g_1$), the $N$-particle quantum Hamiltonian with potential \eqref{Vn}
is quasi-exactly solvable or even, in certain cases, exactly solvable.
This will be proved in the next Section, essentially by showing that
$H_N$ lies in the enveloping algebra of a quasi-exactly solvable Lie
algebra of first-order differential operators equivalent to a certain
standard realization of $\sL(N+1)$ (cf.~Eq.~\eqref{slN} below).
%
%
\section{Algebraization of $H_N$}
\label{alghn}
We shall prove in this Section that the Hamiltonian
\eqref{ham2}--\eqref{Vn} derived in the previous Section is algebraic,
provided that $P(z)$, $\Qt(z)$ and (in certain cases) $g_1$ are chosen
appropriately. In analogy with the one-dimensional case, we shall show
that $H_N(\x)$ is equivalent under a change of variables and a gauge
transformation \eqref{gh} to a gauge Hamiltonian $\Hb_N(\bz)$ which
can be written as a quadratic polynomial in the differential operators
\begin{align}
    \D_k&=\d_{\t_k}\,,\qquad
    \N_{jk}=\t_j\,\d_{\t_k}\,,\qquad
    \U_k=\t_k\left(m-\sum_{i=1}^N \t_i\,\d_{\t_i}\right);\notag\\
\label{slN}
    j,k&=1,2,\dots,N\,,
\end{align}
spanning the Lie algebra $\sL(N+1)$. In the previous formulas $m$ is a
non-negative integer, and we have
denoted by
\begin{equation}
\t_k= \sum_{i_1<i_2< \dots <i_k} z_{i_1}z_{i_2}\cdots z_{i_k}\,,
\qquad 1\le k\le N\,,
\end{equation}
the $k$-th elementary symmetric function.
Since the operators \eqref{slN} preserve the finite-dimensional module
\begin{equation}\label{module}
\M_m = \text{span} \left\{ \t_1^{l_1}\t_2^{l_2} \cdots \t_N^{l_N}
: \sum_{i=1}^N l_i \leq m\right\}
\end{equation}
of polynomials of degree $\leq m$ in the symmetric variables
$\t=(\t_1,\dots,\t_N)$, $\Hb_N(\bz)$ is algebraic, possessing (at
most) $\dim\M_m=\binom{m+N}m$ eigenfunctions
$\chi_k\bigl(\t(\bz)\bigr)$ that are polynomials of degree $\le m$ in
$\t$. It follows from the discussion in the previous Section that
$H_N(\x)$ is also algebraic, and that it admits eigenfunctions of the
form
\begin{equation}
    \psi_k(\x)=\mu(\bz)\,\chi_k\bigl(\t(\bz)\bigr)\,,
    \label{defalg}
\end{equation}
where $\mu(\bz)$ is the gauge factor, whose specific form will be
given below (cf.~\eqref{gfactor}), and $\chi_k\bigl(\t(\bz)\bigr)$ is a polynomial
eigenfunction of $\Hb_N(\bz)$ which can be algebraically computed. For
this reason, we shall henceforth use the term \emph{algebraic
eigenfunction} to refer to an eigenfunction of $H_N$ of the form
\eqref{defalg}.

The generators
\begin{equation}
    \D_k\,,\quad\N_{jk}\,;\qquad 1\le j,k\le N\
    \label{borel}
\end{equation}
span the \emph{Borel
subalgebra} $\borel\subset\sL(N+1)$, which preserves the infinite
sequence of
subspaces $\M_0 \subset \M_1 \subset \M_2\subset \dots$.
Therefore, if $\Hb_N$ is a quadratic combination of the differential
operators \eqref{borel} then the physical
Hamiltonian $H_N$ is \emph{exactly solvable.}

It will also prove convenient in what follows
to define the polynomial subspaces
\begin{equation}
    \L_k=\text{span} \left\{ \t_1^{l_1}\t_2^{l_2} \dots
    \t_N^{l_N}:\sum_{i=1}^N i \,l_i =k\right\}
    \label{Lk}
\end{equation}
of all the symmetric polynomials homogeneous of degree $k$ in the
variables $\bz=(z_1,\dots,z_N)$, and their direct sums
\begin{equation}
    \hat\M_m = \bigoplus_{k=0}^m\,\L_k\,.
    \label{Mt}
\end{equation}
Clearly, $\hat \M_m \subset \M_m$, though in general $\hat\M_m$ need
not be invariant under the action of $H_N$. An important exception
occurs when $H_N$ is exactly solvable; indeed, it will be shown in
Section \ref{EnSp} that in this case $H_N$ preserves the infinite
sequence $\hat\M_0\subset\hat\M_1\subset\hat\M_2\subset\dots$. This
important fact shall be used in Section \ref{EnSp} to derive a formula
for the energy spectrum of all the exactly solvable models we shall
construct, and in Section \ref{exa} to exactly compute some
eigenfunctions of $H_N$ and their corresponding energies for an
arbitrary number of particles $N$.

We shall restrict ourselves in this paper to polynomials $P(z)$ whose
corresponding one-dimensional QES operators \eqref{QESgen2} admit
\emph{normalizable} (i.e, square-integrable) eigenfunctions,
\cite{Artemio93}, leaving the periodic cases for a future work. In
this case, there are $5$ canonical forms for $P(z)$, all of which can
be taken as monic polynomials of
degree not greater than $2$. By \eqref{qtrt}, $\tilde Q(z)$ is
then an arbitrary quadratic polynomial
\begin{equation}\label{Q}
\tilde Q(z) = \ct_+\,z^2 + \ct_0\,z + \ct_-\,.
\end{equation}
Let us define the gauge Hamiltonian $\Hb_N(\bz)$ by Eq.~\eqref{gh},
where the gauge factor
\begin{equation}\label{gfactor}
    \mu(\bz) = \prod_{j<k} z_{jk}^a\cdot \prod_k P_k^\frac b2\,
    \exp\left\{\frac c2\,\int^{z_k}\frac{\Qt}P\right\}
\end{equation}
has been chosen by analogy with the one-dimensional case
(cf.~\cite{Artemio93}), and $a,b,c$ are real parameters. (From now on, all indices in sums and products will
implicitly run from $1$ to $N$, with the restrictions indicated under
the summation or product symbols). Using \eqref{Vn} and dropping an
additive constant we obtain the following explicit formula for
$\Hb_N(\bz)$:
\begin{align}\label{gham}
    \Hb_N(\bz) &= -\sum_k P_k\,\d_{z_k}^2-\sum_k
    \left[\left(b+\frac 12\right)P_k'+c\,\Qt_k\right]\d_{z_k}
   -2a\,\sum_{j\ne k}\frac{P_k}{z_{kj}}\,\d_{z_k}
   +\Vb_N(\bz)\,,
\end{align}
where
\begin{equation}\label{Vn2}
 \Vb_N(\bz) = A_1\,\t_1
   + A_2\, \sum_{j\ne k}\frac{P_k}{z_{kj}^2}
   +\sum_k \frac1{P_k}\left[A_3\,{P_k'}^2+A_4\,\Qt_k\,P_k'
   +A_5\,\Qt_k^2\right],
\end{equation}
and the coefficients $A_i$ are given by
\begin{alignat}{3}
A_1 &= \ct_+\left[(N-1)(g_1-ac)-c\right]\,,&\qquad
A_2 &= g_2-a(a-1)\,,&\qquad
A_3 &=-\frac14\,b\,(b-1)
\,,\notag\\
A_4 &= -\frac c4\,(2b-1)\,,&
A_5 &= \frac
14 \,(1-c^2)\,.
\end{alignat}

We want to find sufficient conditions for $\Hb_N$ to be
expressible as a quadratic combination of the generators
(\ref{slN}). From (\ref{Vn2}) it is clear that one such condition
is $A_2=0$, which yields the following relation between the
coupling constant $g_2$ and the exponent $a$ in the gauge factor
(\ref{gfactor}):
\begin{equation}\label{g2}
g_2 = a (a-1)\,.
\end{equation}
The differential part of $\Hb_N$ equals
\begin{align}\label{diff}
    - m\,c\,\ct_+\,\t_1 +c\,\ct_+\,\U_1&\mod \borel\\ \nonumber
    = - m\,c\,\ct_+\,\t_1 &\mod \sL(N+1)\,.
\end{align}
In Cases 1--3, $P(z)=z^2+\ep$ with $\ep=0,\pm1$, and the last
term in (\ref{Vn2}) can be expressed (up to a constant) as
\begin{equation}\label{Vn3}
A_5\,\ct_+^2\sum_k z_k^2+2\ct_+(A_4+\ct_0 A_5)\,\t_1 +
\sum_k\frac{\rho_1 z_k + \rho_0}{P_k}\,.
\end{equation}
for certain constants $\rho_0$ and $\rho_1$. It is then clear that the
terms proportional to $\t_1$ in (\ref{Vn2}) and (\ref{diff}) must
cancel, and the remaining terms in (\ref{Vn3}) must vanish, giving
rise to the conditions
\begin{align}
&A_5\, \ct_+ = 0\label{a7cp} \\ &\rho_0= \rho_1=0
\label{conds} \\ \label{quant}
 &\ct_+\left[(N-1)(g_1-ac)-\left(b+m+\frac12\right)c\right]=0\,.
\end{align}
The analogous conditions for Cases 4 and 5 (when $P(z)=z$ and
$P(z)=1$, respectively) can be found at the end of this Section, when
we discuss those cases in detail.

In particular, if $\ct_+ =0$ then the ``quantization condition''
(\ref{quant}) is automatically satisfied and the term proportional to
the generator $\U_1$ in \eqref{diff} vanishes (this also holds in
Cases 4 and 5), so that in this case $\Hb_N$ is a quadratic
combination of operators belonging to the Borel subalgebra of
$\sL(N+1)$. Therefore the condition $\ct_+ =0$ ensures the \emph{exact
solvability} of the potential \eqref{Vn}. The specific conditions for
$\Hb_N$ to be algebraic, together with the explicit form of the
potential \eqref{Vn} and the normalizability conditions for the
algebraic eigenfunctions, will be discussed on a case by case basis
below.

All the potentials discussed in what follows diverge when $x_j=x_k$
(cf.~Eq.~\eqref{Vn}). For this reason, we shall consider them to be
defined in the open set
\begin{equation}\label{region1}
 x_N<x_{N-1}<\dots<x_1\,.
\end{equation}
Moreover, the finiteness of the mean kinetic energy of the algebraic
eigenfunctions near the hyperplanes $x_j=x_k$ requires that $a>1/2$,
so that $g_2>-\frac 14$ in all cases. In Cases $2$ and $4$ we shall
impose the additional restriction $x_k> 0$ ($1\le k\le N$) to make
the change of variables \eqref{cofv} one-to-one (\eqref{Vn} is
generically singular at $x_k=0$ in these cases), so that the
potential is defined in the open subset
\begin{equation}\label{region2}
 0<x_N<x_{N-1}<\dots<x_1\,.
\end{equation}
This corresponds to choosing a fundamental chamber for the action
of the Weyl group, \cite{OP}.

The following constants will be used throughout this Section:
\begin{alignat}{2}
    \label{Cpm0}
C_{\pm}&=c\, \ct_{\pm}\,,&\qquad C_{0}&=c\, \ct_{0}\,,\\ B&= b +
\frac12 c\, \ct_0\,,&  B_- &= b + c\,\ct_-\,.
\label{Bpm}
\end{alignat}
\subsection{Case 1.\quad $P_k=z_k^2+1\,,\enspace  z_k = \sinh x_k$.}
It can be shown that a necessary and sufficient condition for all
the wavefunctions in the algebraic sector to be normalizable is
\begin{align}
&\ct_+=0\,, \label{norm11}\\ &a(N-1) + B + m<0\,;\label{norm12}
\end{align}
hence \eqref{a7cp} and (\ref{quant}) are automatically satisfied, and
the potential in this case is \emph{exactly solvable.} Conditions
(\ref{conds}) reduce to
\begin{align}
    \left(B-\frac12\right)^2&=
    \frac14\left(1+C_-^2\right)-\eta_0\,,\label{C11}\\
    C_-\left(B-\frac12\right)&=\eta_1\,,
    \label{C12}
\end{align}
where $\eta_0$ and $\eta_1$ are defined by
\begin{equation}
    \eta_0 = \frac14\left(\ct_-^2-\ct_0^2\right),
    \qquad
    \eta_1 = \frac12\,\ct_0\,\ct_-\,.
    \label{etas1}
\end{equation}
The solution to the previous system
becomes unique after imposing the
normalizability condition (\ref{norm12}), and is given by
\begin{equation}
    B=\frac12\left(1-\sqrt s\right)\,,\qquad
    C_-=-\frac{2\eta_1}{\sqrt s}\,,
    \label{bc1}
\end{equation}
with
\begin{equation}
    \frac s2=\frac14-\eta_0+
    \sqrt{\left(\frac14-\eta_0\right)^2+\eta_1^2}\;.
    \label{t}
\end{equation}
The potential (\ref{Vn}) is in this case
\begin{equation}
    V_N(\x) = \sum_k\left(\eta_0+\eta_1\,\sinh x_k\right)\sech^2 x_k
    +g_2\sum_{j\ne k}\frac{\cosh^2 x_k}{(\sinh x_j-\sinh x_k)^2}\,,
    \label{vn1}
\end{equation}
which, after dropping the constant term $g_2 N(N-1)/2$, can be
expressed in the more familiar form
\begin{multline}
    V_N(\x) = \sum_k\left(\eta_0+\eta_1\,\sinh x_k\right)\sech^2 x_k\\
    +\frac{g_2}4\sum_{j\ne k}\left[
    \csch^2\left(\frac{x_j-x_k}2\right)
    - \sech^2\left(\frac{x_j+x_k}2\right)
    \right].\kern20pt
    \label{vn1OP}
\end{multline}
In the latter formula, $\eta_0$ and $\eta_1$ can be taken as arbitrary
real parameters (cf.~\eqref{etas1}) restricted only by the
normalizability condition (\ref{norm12}), which can be written as
\begin{equation}
    m<\frac12\left(\sqrt s-1\right)-a (N-1)\,,
    \label{intc1sol}
\end{equation}
with $s$ defined by Eq.~\eqref{t}. The exactly solvable potential
(\ref{vn1OP}) will admit square-integrable algebraic eigenfunctions if
the latter inequality can be satisfied by a non-negative integer $m$,
i.e, if the right-hand side is positive, which yields the restriction
\begin{equation}
    s>\left[1+2a(N-1)\right]^2.
    \label{qes1}
\end{equation}
The algebraic eigenfunctions \eqref{defalg} in this case have the form
\begin{equation}
    \psi(\x)=\chi(\t)\cdot
    \prod_{j<k}\left(\cosh\frac{x_j+x_k}2\right)^{\!a}\,
    \left(\sinh\frac{x_j-x_k}2\right)^{\!a}
    \cdot
    \prod_k \sech^{\abs B} x_k\,\e^{\frac{C_-}2\,\arctan(\sinh x_k)}\,,
    \label{algeig1}
\end{equation}
where $a$, $B$, and $C_-$ are given by
Eqs.~\eqref{g2}, \eqref{bc1}, and \eqref{t}. The function $\chi(\t)$ is a
polynomial eigenfunction of $\Hb_N$ of degree $\le m$ in the
symmetric variables $\t_k$, where  $m$ is the largest non-negative
integer satisfying \eqref{intc1sol}.

The potential \eqref{vn1OP} is not, as it stands, one of
Olshanetsky--Perelomov's completely integrable models. If we
perform the imaginary translation $x_k \mapsto x_k + \ii\,\frac{\pi}2$
($1\le k\le N$), then \eqref{vn1OP} becomes
\begin{multline}
    V_N(\x)=\frac14\,\sum_k\left[\frac{\be_+(\be_+-1)}{\sinh^2\frac{x_k}2}
    -\frac{\be_-(\be_--1)}{\cosh^2\frac{x_k}2}\right]\\
    +\frac{g_2}4\sum_{j\ne k}\left[
    \csch^2\left(\frac{x_j+x_k}2\right)
    + \csch^2\left(\frac{x_j-x_k}2\right)
    \right],
    \label{VN1OPc}
\end{multline}
with
\begin{equation}
    \be_\pm=B\mp\ii\,\frac{C_-}2\,,
    \label{bpmc}
\end{equation}
which is a completely integrable potential of $BC_N$ type. Note,
however, that the coefficients $\be_\pm(\be_\pm-1)$ are, generally
speaking, \emph{complex} for real values of $B$ and $C_-$. In fact,
these coefficients can be real only in two cases: i) $C_-=0$, and ii)
$B=1/2$. In the first case, both parameters $\be_\pm=B$ are real and
equal, and \eqref{VN1OPc} reduces to the well-known completely
integrable potential of $C_N$ type
\begin{equation}
    V_N(\x)=B(B-1)\sum_k\csch^2 x_k
    +\frac{g_2}4\sum_{j\ne k}\left[
    \csch^2\left(\frac{x_j+x_k}2\right)
    + \csch^2\left(\frac{x_j-x_k}2\right)
    \right].
    \label{VN1BN}
\end{equation}
Indeed, in this case the algebraic eigenfunctions \eqref{algeig1}
reduce to
\begin{equation}
    \psi(\x)=\chi(\t)\cdot
    \prod_{j<k}\left(\sinh\frac{x_j+x_k}2\,
    \sinh\frac{x_j-x_k}2\right)^{\!a}
    \cdot
    \prod_k \sinh^B x_k\,,
\end{equation}
thus recovering the results in Ref.~\cite{OP}. The second case is not
interesting, since $B=1/2$ implies
\begin{equation}
    \be_\pm(\be_\pm-1)=-\frac14\,(1+C_-^2)\,.
    \label{bpmc2}
\end{equation}
Therefore in this case the Hamiltonian is not self-adjoint, since the
potential behaves as
\begin{equation}
    -\frac14\,(1+C_-^2)\,x_k^{-2}<-\frac14\,x_k^{-2}
    \label{notSA}
\end{equation}
as $x_k\to0$. Thus when $C_-\ne0$, i.e, when $\eta_1\ne0$, the
exactly solvable potential (\ref{vn1OP}) does not seem to be
directly related to a root system.
\subsection{Case 2.\quad $P_k=z_k^2-1\,, \enspace z_k = \cosh x_k$.}
In this case, the normalizability of the algebraic eigenfunctions requires
that either
\begin{align}
&C_+<0\label{norm21}\\
\intertext{or}
&C_+=0 \quad \text{and}
\quad a(N-1) + B+m <0\,,\label{norm22}
\end{align}
which leads to two different subcases, the first one consisting
of QES and the second one of ES potentials. We shall therefore discuss
separately the two cases $C_+<0$ and $C_+=0$.
\subsubsection{Case 2.a.\enspace $C_+<0.$}
Conditions (\ref{a7cp}) and (\ref{conds}) reduce in this case
to
\begin{align}
    c^2&=1 \label{C20}\\
    C_+ + C_- &=0\label{C21}\\
    B(B-1) &=\frac{C_0^2}4,
    \label{C22}
\end{align}
while condition (\ref{quant}) can be expressed as
\begin{equation}
    c(N-1)g_1 = a(N-1)+m+b+\frac12\,.
    \label{a3eq}
\end{equation}
Imposing regularity of the wavefunctions on the hyperplanes
$x_k=0$ forces a unique solution to Eq.~(\ref{C22}), given by
\begin{equation}\label{B2def}
B=\frac12\left( 1 + \sqrt{1+C_0^2}\right).
\end{equation}
Taking into account Eqs.~(\ref{C20})--(\ref{a3eq}), the potential
\eqref{Vn} can be written as
\begin{multline}
    V_N(\x)=\sum_k\left[\frac{C_+^2}4 \sinh^2 x_k+C_+
    \left(a(N-1)+B+m+\frac12\right)\cosh x_k+
    \frac{C_0^2}4\,\csch^2 x_k\right]\\
    +\frac{g_2}4\sum_{j\ne k}\left[
    \csch^2\left(\frac{x_j+x_k}2\right)
    + \csch^2\left(\frac{x_j-x_k}2\right)
    \right].
    \label{VN2c+OP}
\end{multline}
This potential is \emph{quasi-exactly solvable,} and it has $\dim
\M_m=\binom{m+N}m$
algebraic eigenfunctions of the form
\begin{equation}
    \psi(\x)=\chi(\t)\cdot\prod_{j<k}\left(\sinh\frac{x_j+x_k}2\,
    \sinh\frac{x_j-x_k}2\right)^{\!a}\cdot
    \prod_k (\sinh x_k)^B\,\e^{\frac{C_+}2 \cosh x_k}\,,
    \label{eig2ne0}
\end{equation}
where $B$ is given by Eq.~\eqref{B2def}, and $\chi(\t)$ is a
polynomial eigenfunction of $\Hb_N$ of degree $\le m$ in the symmetric
variables $\t_k$. The potential \eqref{VN2c+OP} is regular on the
hyperplanes $x_k=0$ if and only if $C_0=0$. In this case $B=1$, and
therefore the algebraic eigenfunctions \eqref{eig2ne0} are naturally
extended as odd functions to the region
\begin{equation}
    \abs{x_N}<\abs{x_{N-1}}<\dots<\abs{x_1}\,.
    \label{regex2}
\end{equation}

A potential somewhat more general than \eqref{VN2c+OP} has been proved
to be completely integrable in the \emph{classical} case by Inozemtsev
and Meshcheryakov, \cite{Inocencio2}, \cite{InoMesLMP}. Its quantum
integrability has been conjectured by the same authors in
Ref.~\cite{Inocencio}, although, to the best of our knowledge, no
proof of this fact has been given. We have shown in this Section that
the potential \eqref{VN2c+OP}, which belongs to the general class
considered in Ref.~\cite{Inocencio}, is \emph{quasi-exactly solvable}. In
fact, we shall explicitly compute in Section \ref{exa} some
eigenfunctions and eigenvalues of the Hamiltonian with potential
\eqref{VN2c+OP} for $m=1$ and $N=3$ particles.
\subsubsection{Case 2.b. \enspace  $C_+=0.$}
In this case the potential \eqref{Vn}, which is \emph{exactly
solvable}, can be written as
\begin{multline}
    V_N(\x)=\frac14\,\sum_k\left[\eta_+\csch^2\!\left(\frac{x_k}2\right)
    -\eta_-\sech^2\!\left(\frac{x_k}2\right)\right]\\
+\frac{g_2}4\sum_{j\ne k}\left[
    \csch^2\left(\frac{x_j+x_k}2\right)
    + \csch^2\left(\frac{x_j-x_k}2\right)
    \right],
    \label{VN20OP}
\end{multline}
where
\begin{equation}
    \eta_\pm=\frac14\,(\ct_-\pm\ct_0)^2\ge0\,.
    \label{eta01}
\end{equation}
Conditions (\ref{a7cp}) and (\ref{quant}) are now
automatically satisfied, while (\ref{conds}) reduces to
\begin{align}
    \left(B-\frac12\right)^2&=
    \frac14\left(1-C_-^2\right)+\frac12\,(\eta_++\eta_-)\label{C2b1}\\
    C_-\left(B-\frac12\right)&=\frac12\,(\eta_+-\eta_-)\,.
    \label{C2b2}
\end{align}
Introducing the parameters
\begin{equation}
    \beta_\pm=B\pm\frac{C_-}2\,,
    \label{betapm}
\end{equation}
the latter system reduces to
\begin{equation}
    \be_\pm(\be_\pm-1)=\eta_\pm\,,
    \label{bpmcond}
\end{equation}
which uniquely determines $\be_\pm$. Indeed, the algebraic
eigenfunctions in this case are of the form
\begin{equation}
    \psi(\x)=\chi(\t)\cdot\prod_{j<k}\left(\sinh\frac{x_j+x_k}2\,
    \sinh\frac{x_j-x_k}2\right)^{\!a}\cdot
    \prod_k \sinh^{\beta_+}\!\left(\frac{x_k}2\right)\,
    \cosh^{\beta_-}\!\left(\frac{x_k}2\right),
    \label{psi20}
\end{equation}
where $\chi(\t)$ is a polynomial eigenfunction of $\overline H_N$ of
degree $\le m$ in the symmetric variables $\t$.
Therefore $\be_+\ge1/2$ for the eigenfunctions to be regular (with
finite kinetic energy) on the hyperplanes $x_k=0$,
while $\beta_-\le-1/2$ is necessary for the normalizability condition
\eqref{norm22}, which now reads
\begin{equation}
    \frac12\,(\be_++\be_-)+a(N-1)+m<0\,,
    \label{norm2es}
\end{equation}
to hold. Thus Eqs.~\eqref{bpmcond} have the unique solution
\begin{equation}
    \be_\pm=\frac12\left(1\pm\sqrt{1+4\,\eta_\pm}\right),
    \label{bpmsol}
\end{equation}
from which it actually follows that $\be_+\ge1$ and $\be_-\le-1$.

The exactly solvable potential \eqref{VN20OP} is known to be
integrable, \cite{OP}, and is associated with the root system $BC_N$,
when $\eta_+\ne\eta_-$, or $C_N$, when $\eta_+=\eta_-$. The potential
is regular on the hyperplanes $x_k=0$ if and only if $\be_+=1$, in
which case the algebraic eigenfunctions \eqref{psi20} are naturally
extended as odd functions to the region \eqref{regex2}.
\subsection{Case 3.\quad  $P_k=z_k^2\,,\enspace z_k = \e^{x_k}$.}
For the algebraic eigenfunctions to be square-integrable, one of the
following three conditions must be satisfied:
\begin{enumerate}
\item $C_+<0 \,,\enspace C_->0$\label{norm31}
\item $C_+=0\,,\enspace C_->0\,,\enspace a(N-1) + B+m <0$\label{norm32}
\item  $C_+<0\,,\enspace C_-=0\,,\enspace B>0$\label{norm33}
\end{enumerate}
In principle, the first case yields a family of QES potentials
and the remaining two cases a family of ES ones. However, it can be
shown that the potential in the third case can be
obtained from that in the second one by the reflection
$\x\mapsto -\x$. We shall therefore consider only the first two cases.
\subsubsection{Case 3a.\enspace $C_+<0\,,\enspace C_->0.$}
Conditions (\ref{a7cp}) and (\ref{conds}) are satisfied
if
\begin{equation}
    \label{c2b}
c^2=1\,, \quad b=\frac12\,,
\end{equation}
and the quantization condition (\ref{quant}) can be written
as
\begin{equation}
    c\,g_1\,(N-1)=a(N-1)+b+m+\frac12\,.
    \label{g13}
\end{equation}
The potential (\ref{Vn}) becomes in this case
\begin{align}
    V_N(\x)=&\sum_k\left[\frac{C_-^2}4\,\e^{-2x_k}+
    \frac12 C_0\,C_-\e^{-x_k}+
    C_+\bigl(a(N-1)+\frac{C_0}2+m+1\bigr)\e^{x_k}+
    \frac{C_+^2}4\,\e^{2x_k}\right]
    \notag\\
    &\quad{}+\frac{g_2}4\sum_{j\ne k}\csch^2\left(\frac{x_j-x_k}2\right).
    \label{vn3q1}
\end{align}
The quantum Hamiltonian (\ref{ham2}) with potential (\ref{vn3q1})
defined in the region \eqref{region1} is \emph{quasi-exactly
solvable.} It possesses $\binom{m+N}m$ algebraic eigenfunctions of
the form
\begin{equation}
    \psi(\x)=\chi(\t)\cdot\prod_{j<k}\left(\e^{x_j}-\e^{x_k}\right)^a\,
    \cdot \prod_k
    \e^{\frac12(C_0+1)x_k}\,\e^{\frac12\left(C_+\e^{x_k}-C_-\e^{-x_k}
    \right)},
    \label{eig3q1}
\end{equation}
which can be more conveniently written as
\begin{equation}
    \psi(\x)=\chi(\t)\cdot\prod_{j<k}\sinh^{\,a}\left(\frac{x_j-x_k}2
    \right)\,
    \cdot \prod_k
    \e^{\frac12[C_0 + a(N-1)+1]x_k}\,
    \e^{\frac12\left(C_+\e^{x_k}-C_-\e^{-x_k}\right)},
    \label{eig3q2}
\end{equation}
where $\chi(\t)$ is a polynomial eigenfunction of $\Hb_N$ of
degree $\le m$ in the symmetric variables $\t_k$.

The potential \eqref{vn3q1} with arbitrary constants multiplying each
exponential has been shown to be classically completely integrable in
Ref.~\cite{Inocencio2}, but the explicit solutions of the equations of
motion could only be calculated when certain restrictions on the
constants were imposed, and only for a subset of initial conditions,
\cite{Inocencio3}. In the quantum case, no exact formulas for the
eigenfunctions and their corresponding energies are available when
both the negative and positive exponentials are present, although
these models have been conjectured to be completely integrable also in
this case, \cite{Inocencio}. We have proved that for certain values of
the coefficients multiplying the exponentials the potential
\eqref{vn3q1} is \emph{quasi-exactly solvable,} and thus some of its
energies and eigenfunctions (in general corresponding to excited
states) can be exactly computed. See Section \ref{exa} for a concrete
example when $m=1$.

\subsubsection{Case 3b.\enspace $C_+ =0\,,\enspace C_->0.$}
The conditions (\ref{a7cp}) and (\ref{conds}) lead again to \eqref{c2b},
but now the quantization condition (\ref{quant}) is absent.
The potential can be simply obtained by making $C_+=0$ in
\eqref{vn3q1}:
\begin{equation}
    V_N(\x)=\sum_k\left[\frac{C_-^2}4\,\e^{-2x_k}+
    \frac12 C_0\,C_-\e^{-x_k}
    \right] +\frac{g_2}4\sum_{j\ne
    k}\csch^2\left(\frac{x_j-x_k}2\right).
    \label{vn3es}
\end{equation}
Note that this potential depends effectively only on one parameter,
since the ratio $\abs{C_-/C_0}$ can be assigned any prescribed
positive value by performing a suitable translation of the
coordinates. The same argument shows that the discrete spectrum of the
potential \eqref{vn3es} cannot depend on $C_-$.

The quantum Hamiltonian (\ref{ham2}) with potential (\ref{vn3es}) is
\emph{exactly solvable}. It describes an external Morse potential
acting on each particle, plus an $A_N$-type hyperbolic interaction.
The discrete spectrum of this potential has been computed by
Inozemtsev and Meshcheryakov, \cite{Inocencio}, by relating
\eqref{vn3es} to the integrable $BC_N$ potential \eqref{VN20OP} through
a certain formal limit.

The square-integrability of the algebraic eigenfunctions is ensured if
Eq.~\eqref{norm12} holds or, taking into account \eqref{c2b},
\begin{equation}
    C_0+2a(N-1)+2m+1<0\,.
    \label{norm3b}
\end{equation}
This inequality can be satisfied for non-negative integer values of $m$
provided that
\begin{equation}
    C_0+2a(N-1)+1<0\,.
    \label{norm3b0}
\end{equation}
In particular, note that the latter inequality implies that $C_0$ is
negative, and therefore the coefficient of $e^{-x_k}$ in the potential
is negative as well.
The algebraic eigenfunctions are given in this case by
\begin{equation}
    \psi(\x)=\chi(\t)\cdot\prod_{j<k}\sinh^{\,a}\left(\frac{x_j-x_k}2
    \right)\,\cdot \prod_k
    \exp\left(\frac12[C_0 + a(N-1) +1]x_k -\frac{C_-}2\e^{-x_k}\right),
    \label{alg0}
\end{equation}%
where $\chi(\t)$ is a polynomial eigenfunction of $\Hb_N$ of degree
$\le m$ in $\t$, $m$ being the largest non-negative integer compatible
with \eqref{norm3b}.

The limit $C_-=0$ corresponds to one of Olshanetsky--Perelomov's
integrable potentials associated to the $A_N$ root system, namely
\begin{equation}
    V_N(\x)=\frac14a(a-1)\sum_{j\ne
k}\csch^2\left(\frac{x_j-x_k}2\right),
    \label{vn3OP}
\end{equation}
whose algebraic eigenfunctions are no longer normalizable (indeed,
this potential is known to have no bound states, \cite{OP}).

\subsection{Case 4.\quad $P_k=z_k\,, \enspace z_k = \ds\frac{x_k^2}4$.}
The algebraic eigenfunctions will be square integrable provided
that
\begin{align}
&C_+<0\,\label{norm41}\\
\intertext{or}
&C_+=0\,, \quad C_0 <0\,.\label{norm42}
\end{align}
Conditions (\ref{a7cp})--(\ref{quant}) must be modified in this case
(since $P$ is no longer quadratic), but after applying similar
considerations it can be shown that the operator $\Hb_N$ is algebraic
provided that
\begin{align}
&c^2 =1 \\ & B_-(B_- -1) = C_-^2\label{b4}\\
&C_+\left[(a-c\,g_1)(N-1)+m+\frac b2+\frac34\right]=0\,.
\label{g1n1}
\end{align}
The potential is given by
\begin{equation}
    V_N(\x) = \sum_k \left[
    \eta_0^2\,x_k^6+\eta_1\,x_k^4+\eta_2\,x_k^2+\frac{\eta_3}{x_k^2}
    \right]
    +g_2\sum_{j\ne k}\left[\frac1{(x_j+x_k)^2}+\frac1{(x_j-x_k)^2}
    \right]\,,
    \label{vn4qes}
\end{equation}
where
\begin{align}
    \eta_0&=-\frac{C_+}{16}\,,\qquad
    \eta_1=\frac{C_0 C_+}{32}\,,\notag\\
    \eta_2&=\frac{C_0^2}{16}+\frac{C_+}4\left[
    a(N-1)+m+\frac {B_-}2 +\frac34\right]\,,\\
    \eta_3&=C_-^2\ge0\,,\notag
\end{align}
and
\begin{equation}
    B_-=\frac12\left(1+\sqrt{1+4\,\eta_3}\right).
    \label{B4}
\end{equation}
(We have taken the positive square root to ensure that the algebraic
eigenfunctions are regular when $x_k \rightarrow 0$.)
When $C_+\ne 0$ the potential (\ref{vn4qes}) is
\emph{quasi-exactly solvable}. The coefficients in the
potential are not all independent in this case, but satisfy the relation
\begin{equation}
   \eta_2=\frac{\eta_1^2}{4\eta_0^2}-4 \eta_0\left[
    a(N-1)+m+ \frac{B_-}2+\frac34\right].
    \label{eta24}
\end{equation}
The algebraic eigenfunctions are of the form
\begin{equation}
    \psi(\x)=\chi(\tau)\cdot\prod_{j<k}\left(x_j^2-x_k^2\right)^a\cdot
    \prod_k x_k^{B_-}
    \e^{-\frac1{4\eta_0}\left(\eta_0^2\,x_k^4
    +\eta_1\,x_k^2\right)}\,,
    \label{eig4qes}
\end{equation}
where $B_-$ is given by \eqref{B4}, and $\chi(\t)$ is a polynomial
eigenfunction of $\Hb_N$ of degree $\le m$ in $\t$. This QES potential
has been recently obtained in somewhat less generality by Hou and
Shifman, \cite{Shifman}.

If $C_+=0$ then the potential \eqref{vn4qes} becomes \emph{exactly
solvable}, and assumes the simpler form
\begin{equation}
    V_N(\x)=\sum_k\left(\frac{\omega^2}4\,x_k^2+\frac{\eta_3}{x_k^2}\right)
    +g_2\sum_{j\ne
    k}\left[\frac1{(x_j+x_k)^2}+\frac1{(x_j-x_k)^2}\right],
    \label{vn4es}
\end{equation}
where now the parameters $\omega\equiv 2\sqrt {\eta_2}=\abs{C_0}/2>0$ and
$\eta_3\ge0$ are independent. The algebraic eigenfunctions are in
this case
\begin{equation}
        \psi(\x)=\chi(\tau)\cdot\prod_{j<k}\left(x_j^2-x_k^2\right)^a\cdot
    \prod_k x_k^{B_-}
    \e^{-\frac{\omega}4\,x_k^2}\,,
    \label{eig4es}
\end{equation}
where $B_-$ and $\chi(\t)$ are as before. This potential is well
known to be integrable, \cite{OP}, and is commonly known as a rational
$B_N$-type potential with harmonic force.

The potentials \eqref{vn4qes}--\eqref{vn4es} are regular on the
hyperplanes $x_k=0$ provided that $\eta_3=0$, or, equivalently,
$B_-=1$. By \eqref{eig4qes}--\eqref{eig4es}, in this case the algebraic
eigenfunctions can again be extended as odd eigenfunctions to the
region \eqref{regex2}.

Following the idea of Minzoni, Rosenbaum and Turbiner,
\cite{Turb2}, we can obtain a different QES deformation of
the ES potential \eqref{vn4es}. Note that the gauge hamiltonian $\Hb_N$
in this case can be written as
$$
\Hb_N = h^{(1)} + h^{(2)}\,,
$$
where
\begin{align}
 &h^{(1)}=-\t_1\,\d_{\t_1}^2-(A
   +C_0\,\t_1)\,\d_{\t_1}\,,
   \label{hexp} \\
&      A\equiv N\left[B_-+\frac12+a(N-1)\right],
   \label{Aeq}
\end{align}
depends only on $\t_1$, and
$$
h^{(2)}\cdot\phi(\t_1) = 0\,.
$$
If we consider eigenfunctions $\chi$
of $\Hb_N$ depending only on $\t_1$ then the problem reduces to
the one-dimensional equation
\begin{equation}
-\t_1 \chi'' - (A+ C_0 \t_1) \chi' =E \chi\,.
\end{equation}
If we add a function $U(\t_1)$ to the potential \eqref{vn4es}, we can
still look for eigenfunctions of $\Hb_N$ depending only on $\t_1$,
which must satisfy the equation
\begin{equation}
    \label{modeig}
\t_1 \chi'' + (A+ C_0 \t_1) \chi' + [E-U(\t_1)] \chi=0\,.
\end{equation}
A wide class of exact solutions of \eqref{modeig} is obtained when the
operator $h= h^{(1)} + U(\t_1)$ is equivalent under a gauge
transformation to a quasi-exactly solvable operator on the line. The
most general such operator equivalent to $h$ must be of the form
\begin{equation}
   \bar h = -\t_1\,\d_{\t_1}^2+(2\ga\,\t_1^2+\be\,\t_1+\a)\,\d_{\t_1}
   -2n\ga\,\t_1+E_0\,,
   \label{hh}
\end{equation}
where $E_0$ is a constant and $n$ is a non-negative integer,
\cite{Artemio93}. The latter operator leaves invariant the space of
polynomials in $\t_1$ of degree not greater than $n$, and consequently
admits $n+1$ algebraic eigenfunctions. For $h$ and
$\bar h$ to be equivalent we must have
\begin{equation}
   \bar h = \e^{-f(\t_1)}\cdot h\cdot \e^{f(\t_1)}\,.
   \label{conj}
\end{equation}
From this relation we obtain the equations
\begin{align}
   2\t_1\,f'+A+C_0\,\t_1&=-\left(2\ga\,\t_1^2+\be\,\t_1+\a\right)\\
   \t_1(f''+f'{}^2)+(A+C_0\,\t_1)f'-U&=2n\ga\,\t_1-E_0\,,
\end{align}
from which we easily get
\begin{equation}
   f=-\frac12\left[\ga\,\t_1^2+(\be+C_0)\t_1+(A+\a)\log\t_1\right]
   \label{f}
\end{equation}
and (after dropping a constant term)
\begin{equation}
   U(\t_1) =
\ga^2\,\t_1^3+\be\ga\,\t_1^2+\frac14\left[\be^2
-C_0^2+4\ga(\a-2n-1)\right]\t _1
-\frac{(A-\a-2)(A+\a)}{4\,\t_1}\,.
   \label{U}
\end{equation}
If we define
$$
r^2= \sum_k x_k^2= 4\,\t_1\,,
$$
then we have shown that the $N$-particle potential
\begin{align}
\tilde V_N(\x) &= V_N(\x) + U(r^2/4)\notag \\ &= \tilde A\,r^6
+ \tilde B\,r^4 + \tilde C\,r^2 + \frac{\tilde D}{r^2} +
B_-(B_--1)\sum_k \frac1{x_k^2}  \nonumber\\ &\quad{}+ a(a-1)\sum_{j\ne
k}\left[\frac1{(x_j+x_k)^2}
   +\frac1{(x_j-x_k)^2}\right],
\end{align}
with
\begin{alignat}{2}
\tilde A &= \frac{\ga^2}{64}\,, &\qquad \tilde
C&=\frac1{16}\left[\be^2+ 4 \ga(\alpha-2n-1)\right],\\ \tilde B &=
\frac{\be\ga}{16}\,, &\tilde D &= -(A-\alpha-2)(A+ \alpha),
\end{alignat}
possesses $n+1$ algebraic eigenfunctions of the form
\begin{align}
   \psi_k(\x)=
   r^{-(A+\a)}\,\exp\left(-\frac\ga{32}\,r^4-\frac{\be}8\,
   r^2\right)\,\sigma_k(r^2/4)\cdot\prod_{j<k}(x_j^2-x_k^2)^a\cdot\prod_k
   x_k^{B_-}\,,
   \label{eigun}
\end{align}
where $\sigma_k(\t_1)$ is a polynomial eigenfunction of $\bar h$ of degree
less than or equal to $n$. These eigenfunctions are
square-integrable and regular at the origin provided that
\begin{equation}
    \ga>0\quad\text{(or $\ga=0$ and $\be>0$)},\qquad
    \a>\frac N2\,,\qquad B_->\frac12\,.
    \label{params}
\end{equation}
\subsection{Case 5. \quad $P_k=1\,,\enspace z_k = x_k$.}
The conditions for $\Hb_N$ to be algebraic reduce in this case to
\begin{equation}
A_5\,\ct_+=A_5\,\ct_0=0\,.
\label{conds5}
\end{equation}
On the other hand, the necessary and sufficient
conditions for the algebraic eigenfunctions to be normalizable are
\begin{equation}
C_+=0\,,\quad C_0<0\,.
\label{norm5}
\end{equation}
Hence the all potentials in this class are \emph{exactly solvable.}
From Eqs.~\eqref{conds5} and \eqref{norm5} it follows that
$c^2=1$, so that the potential \eqref{Vn2} reduces to the celebrated
Calogero model
\begin{equation}
    V_N(\x)=\o^2\,\sum_k x_k^2+g_2\sum_{j\ne
    k}\frac1{(x_j-x_k)^2}\,,
    \label{v5}
\end{equation}
with
\begin{equation}
    \o = \frac{\abs{C_0}}2>0\,.
    \label{eta05}
\end{equation}
(To obtain the previous formula, we have performed a constant 
translation to get rid of the irrelevant parameter $C_-$.)
The algebraic eigenfunctions are of the form
\begin{equation}
    \psi(\x)=\chi\bigl(\t(\x)\bigr)\cdot\prod_{j<k}(x_j-x_k)^a\cdot
    \prod_k \e^{-\frac\o2\,x_k^2}\,,
    \label{eig5}
\end{equation}
where the functions $\chi(\t)$ are polynomial eigenfunctions of
$\Hb_N$, known in the literature as the \emph{Calogero
polynomials}.
%
%
\section{Energy Spectrum}
\label{EnSp}
In this Section we shall obtain an explicit expression for the energy
spectrum of all the ES models derived in the previous Section,
discussing some of its basic properties. The key idea in this respect
is to show that the ES Hamiltonians preserve not only the subspaces
$\M_k$, but also the smaller subspaces $\hat \M_k$. We shall then
define a lexicographic ordering of the basis elements within each
subspace $\hat \M_k$ ensuring that the matrix of the restriction
of $\Hb_N$ to $\hat \M_k$ is triangular. The spectrum of $\Hb_k$, and
therefore that of $H_N$, will thus simply consist of the diagonal
elements of these matrices for $k=0,1,2,\dots$.
\subsection{Cases 1--3}

The gauge Hamiltonian \eqref{gham} corresponding to the ES
Hamiltonians with potential (\ref{vn1OP}), (\ref{VN20OP}) and
(\ref{vn3es}) can be written as
\begin{equation}
    \Hb_N(\bz) =E_0 -\sum_k (z_k^2+\epsilon)\,\d_{z_k}^2-\sum_k
\left[(2B +1) z_k + C_- \right]\d_{z_k}
   -2a\,\sum_{j\ne k}\frac{z_k^2+\epsilon}{z_{kj}}\,\d_{z_k}\,,
\end{equation}
where $\epsilon=1,-1,0$, respectively, and
\begin{equation}\label{gstate}
 E_0=-N\left\{ \left[B+ \frac a2(N-1) \right]^2 +
 \frac{a^2}{12} (N^2-1) \right\}.
\end{equation}
For the potentials \eqref{vn1OP} and \eqref{VN20OP} $B$ and $C_-$ are
given by Eqs.~\eqref{bc1}--\eqref{t} and \eqref{betapm}, respectively,
while $B=\frac12(C_0+1)$ for the potential \eqref{vn3es} on account of
\eqref{c2b}. The expression of $\Hb_N$ in terms of the symmetric
variables is easily found to be
\begin{align}\label{ghamtau}
\Hb_N = E_0&-\sum_{i,j}A_{ij}^{(2)} \d_{\t_i} \d_{\t_j} -
(2B+1)\sum_j j \t_j \d_{\t_j} - a \sum_j j (2N-j-1) \t_j \d_{\t_j}
\notag\\  &{}- C_- \sum_j (N-j+1) \t_{j-1} \d_{\t_j} \\ \notag &
{}+\epsilon \left( -\sum_{ij} A_{ij}^{(0)}\d_{\t_i} \d_{\t_j} + a
\sum_j  (N-j+1)(N-j+2) \t_{j-2} \d_{\t_j}\right),
\end{align}
where the coefficients $A_{ij}^{(p)}$ can be found in the Appendix.

It follows from the latter expression and the structure of the
coefficients $A_{ij}^{(p)}$ that $\Hb_N$ preserves the subspaces
$$
    \hat\M_k = \bigoplus_{j=0}^k\,\L_j\,
$$
for arbitrary $k$, where
$$
    \L_j=\text{span} \left\{ \t_1^{l_1}\t_2^{l_2} \dots
    \t_N^{l_N}:\sum_{i=1}^N i \,l_i =j\right\}\,.
$$
In fact, the first line of \eqref{ghamtau} leaves the subspace
$\L_k$ invariant, while the second and third lines take an element
of $\L_k$ into $\L_{k-1}$ and $\L_{k-2}$, respectively.

We now introduce an ordering of the basis of $\hat
\M_k$ consisting of monomials. First of all, we declare the monomials
belonging to $\L_j$ to be less than those in $\L_k$ if $j<k$. Within each
$\L_j$, the ordering is then defined as follows: if the
monomial $\t_1^{l_1}\dots\t_N^{l_N}$ is denoted by the multi-index
$l\equiv(l_1,l_2,\dots,l_N)$, then
\begin{equation}\label{order}
l < l' \quad \text{if} \quad l_N=l'_N,\;l_{N-1} = l'_{N-1},\;
\dots\;,\; l_{i+1}=l_{i+1}',\;l_i > l'_i\,.
\end{equation}
Thus, for instance, $\t_j<\t_1 \t_{j-1} < \ldots <\t_1^{j-2}\t_2<\t_1^j.$

Given this ordering of the basis, it is straightforward to check that
the matrix $\Hb_N |_{\hat \M_k}$ is upper triangular. Its diagonal
elements
\begin{equation}
    \label{eiges}
E_{l_1\,l_2\ldots l_N}= E_0- \left( \sum_i i\,l_i \right)\left[2B +
a(2N-1)+ \sum_i l_i\right] + a \sum_i i^2\,l_i
\end{equation}
give therefore the energy spectrum of $\Hb_N$, and hence of $H_N$.
The spectrum consists of a term
$$
E_0-\left[2B+a(2N-1)\right]\sum_i i\,l_i
$$
which is constant over $\L_j$, while the remaining two terms split the
energies within this subspace, although in general they do not remove
all degeneracy. It can be proved with the aid of the normalizability
condition that the lowest energy corresponding to each block $\L_j$ is
$E_{j0\dots0}$ while the highest one is $E_{0\dots 1\dots r}$, where
the $1$ is in the $p$-th position, and $j=N\,r+p$ with $1\le p\le
N-1$. This easily implies that the lowest energy in each block $\L_j$
is lower than the lowest energy in the next block $\L_{j+1}$. In
particular, the first excited state always lies in $\hat\M_1$ (since
$\dim\M_0=1)$, and can thus be exactly computed for any number of
particles. On the other hand, some energies in $\L_{j+1}$ might be
lower than those of $\L_j$. Note also that
the formula \eqref{eiges} for the eigenvalues of the Hamiltonian with
potential \eqref{VN20OP} and \eqref{vn3es} is considerably simpler
than the one given in Ref.~\cite{Inocencio}.

The spectrum of the three ES potentials (\ref{vn1OP}), (\ref{VN20OP})
and (\ref{vn3es}) is the same when expressed in terms of $B$ and
$C_-$, although $B$ and $C_-$ are differently related to the parameters
appearing in these potentials. Due to the triangular nature of the
matrix representing $\Hb_N$ in $\hat\M_k$, it should not be difficult
to give recursive expressions for the eigenfunctions, though we
postpone this task for a future work.

\subsection{Cases 4--5}

The spectrum of the potentials \eqref{vn4es} and \eqref{v5} is highly
degenerate, since in this case the energy $E_{l_1\dots l_N}$ is easily
seen to depend only on the single quantum number $j=\sum_{i=1}^N
i\,l_i$. More precisely, using the explicit expression \eqref{gham} for
the gauge Hamiltonian and passing to the symmetric variables it is 
straightforward to show that
\begin{equation}
E_j= E_0 + 2 \omega j\,, \qquad j=0,1,2,\dots\,,
\end{equation}
where the ground state energy $E_0$ is given respectively by
\begin{align}
E_0^{(4)}=& N[1 + 2B_- + 2a(N-1)]\,\frac\omega2\,,\\
E_0^{(5)}=&N[ 1 + a(N-1)]\,\omega\,.
\end{align}
We have thus rederived (albeit making no use of the underlying root 
systems) the well-known result that the energy levels 
in these models are equally spaced, \cite{OP}.
%
%
\section{Examples}
\label{exa}
\subsection{ES potential from Case 1}
For the ES potential \eqref{vn1OP}
\begin{multline}
    V_N(x) = \sum_k\left(\eta_0+\eta_1\,\sinh x_k\right)\sech^2 x_k\\
    +\frac{g_2}4\sum_{j\ne k}\left[
    \csch^2\left(\frac{x_j-x_k}2\right)
    - \sech^2\left(\frac{x_j+x_k}2\right)
    \right]\kern20pt
    \label{vn1OPex}
\end{multline}
the gauge Hamiltonian in the symmetric variables is explicitly given by
\begin{align}
    \Hb_N = &-\sum_{i,j}\left[j\,\t_i\t_j+\sum_{k=1}^j
    (j-i-2k)\t_{i+k}\t_{j-k}\right]\d_{\t_i}\d_{\t_j}\notag\\
    &{}-\sum_{i,j}\left[(N-i+1)\t_{i-1}\t_{j-1}
   -\sum_{k=1}^{j-1}(i-j+2k)\t_{i+k-1}\t_{j-k-1}\right]\d_{\t_i}\d_{\t_j}
   \notag\\
    &{}-\sum_j j\left[2B+1+a(2N-j-1)\right]\t_j\d_{\t_j}
    -C_-\sum_j(N-j+1)\t_{j-1}\d_{\t_j}\notag\\
    &{}+a\sum_j(N-j+1)(N-j+2)\t_{j-2}\,\d_{\t_j}
    +E_0\,,
    \label{hbnex1}
\end{align}
where $B$, $C_-$, and $E_0$ are given by Eqs.~\eqref{bc1}--\eqref{t}
and \eqref{gstate}, respectively. We know from the previous Section
that the gauge Hamiltonian \eqref{hbnex1} preserves each
subspace $\hat{\mathcal M}_k$ for arbitrary $k$. For instance, the
restriction of $\Hb_N$ to $\hat{\mathcal M}_2$ has the following
matrix with respect to the basis $\left\{1,\t_1,\t_2,\t_1^2\right\}$
of $\hat{\mathcal M}_2$:
\begin{equation}
    \begin{pmatrix}
    E_0 & -NC_- & aN(N-1) & -2N\\
    0 & E_0+1-\be & -(N-1)C_- & -2NC_-\\
    0 & 0 & E_0+2(a-\be+1) & 4\\
    0 & 0 & 0 & E_0-2\be
    \end{pmatrix},
    \label{hb1mat}
\end{equation}
where we have set
\begin{equation}
    \frac\be2=a(N-1)+B+1<-1\,.
    \label{betaex}
\end{equation}
(The latter inequality is obtained by imposing the normalizability of
the algebraic eigenfunctions belonging to $\hat\M_2$.) We can exactly
compute four eigenfunctions of the potential \eqref{vn1OPex} with
their corresponding energies by diagonalizing the matrix
\eqref{hb1mat}. The energies are simply the eigenvalues of this
matrix, namely its diagonal elements
\begin{align*}
    &E_0\,,\\
    & E_1=E_0-\be+1\,,\\
    & E_2=E_0-2\be\,,\\
    & E_3 = E_0+2(a-\be+1)\,.
    \label{en1}
\end{align*}
The corresponding eigenfunctions are given by
\begin{equation}
    \psi_i=\mu\,\chi_i\,,\qquad 0\le i\le 3\,,
    \label{eigex1}
\end{equation}
where
\begin{equation}
    \mu=\prod_{j<k}\left(\cosh\frac{x_j+x_k}2\,
    \sinh\frac{x_j-x_k}2\right)^{\!a}\cdot
    \prod_k (\sech x_k)^{\abs B}\,\e^{\frac{C_-}2\,\arctan(\sinh x_k)}
    \label{muex1}
\end{equation}
and
\begin{align}
    \chi_0&=1\\
    \chi_1&=\t_1+\frac{N C_-}{\be-1}\\
    \chi_2&=\t_1^2-\frac{2\t_2}{a+1}
    +\frac{2C_-(Na+1)}{(a+1)(\be+1)}\,\t_1
    +\frac{N(aN+1)(C_-^2+\be+1)}{\be(a+1)(\be+1)}\\
    \chi_3&=\t_2+\frac{C_-(1-N)}{2a-\be+1}\t_1
    +\frac{N(N-1)(2a^2-a\be+a+C_-^2)}{2(a-\be+1)(2a-\be+1)}\,,
\end{align}
with
\begin{equation}
    \t_1=\sum_k \sinh x_k\,,\qquad
    \t_2=\sum_{j<k}\sinh x_j\,\sinh x_k\,.
    \label{t12}
\end{equation}%
Note that, since $\psi_0=\mu$ has no zeros in the configuration space
\eqref{region1}, it is the ground state of the system. We have thus
been able to exactly compute some eigenvalues and eigenfunctions of
the potential \eqref{vn1OPex} for an \emph{arbitrary} number of
particles $N$.
\subsection{QES potential from Case 2}
For simplicity, we shall restrict the number of particles to $N=3$
and set $m=1$ to find $4$ eigenvalues of the potential
\eqref{VN2c+OP}
\begin{multline}\label{QESex}
    V_3(\x)=
    \sum_{k=1}^3\left[B(B-1)\csch^2 x_k+\frac{C_+^2}4 \sinh^2 x_k+C_+
    \left(2a+B+\frac32\right)\cosh x_k\right]\\
+\frac{a(a-1)}4\, \sum_{\substack{j,k=1\\j\ne k}}^3 \left[
    \csch^2\left(\frac{x_j+x_k}2\right)
    + \csch^2\left(\frac{x_j-x_k}2\right)
    \right],
\end{multline}
which can be thought of as a deformation of a $C_N$-type
hyperbolic potential. The gauge hamiltonian $\Hb_3 = \mu^{-1}\,
H_3\, \mu $ can be written as
$$
-\sum_{k=1}^3 (z_k^2-1)
\d_{z_k}^2 -\sum_{k=1}^3 [(2 B+1) z_k + C_+(z_k^2-1)]\,\d_{z_k} -
2a \sum_{\substack{j,k=1\\j\neq k}}^3\frac{z_k^2-1}{z_k-z_j}\,\d_{z_k} +
\,C_+\sum_{k=1}^3 z_k+\ep_0\,,
$$
where $z_k = \cosh x_k$,
\begin{equation}
    \mu=\prod_{1\le j<k\le 3}\left(\sinh\frac{x_j+x_k}2\,
        \sinh\frac{x_j-x_k}2\right)^{\!a}\cdot
        \prod_{k=1}^3 (\sinh x_k)^B\,\e^{\frac{C_+}2 \cosh x_k}\,,
    \label{muqesex}
\end{equation}
and
\begin{equation}
    \ep_0=-\left(3B^2+6aB+5a^2\right).
    \label{E03}
\end{equation}
When expressed in terms of the symmetric variables, the operator
$\Hb_3$ has been shown in Section \ref{alghn} to leave invariant the
subspace
$$
\M_1 =
\text{span} \left\{ 1,\t_1, \t_2, \t_3 \right\}.
$$
The restriction of $\Hb_3$ to this subspace is explicitly given by
\begin{align*}
 \Hb_3\mid_{\M_1} =& -\sum_{j=1}^3 [2B+1+a(5-j)]\,j\,\t_j\, \d_{\t_j} -
 C_+ \sum_{j=1}^3[\t_1 \t_j - (j+1) \t_{j+1}] \,\d_{\t_j} \\
 &+ C_+ \sum_{j=1}^3 (4-j) \t_{j-1}\, \d_{\t_j} - a \sum_{j=1}^3
 (5-j)(4-j) \t_{j-2} \,\d_{\t_j} + C_+ \t_1+\ep_0\,.
\end{align*}
The spectral problem for $\Hb_3$ in the subspace $\M_1$ thus reduces to
diagonalizing the $4\times4$ matrix of  $\Hb_3|_{\M_1}$  with respect
to the basis $\{1,\t_1,\t_2,\t_3\}$, given by
\begin{displaymath}
\begin{pmatrix}\ep_0 & 3 C_+ & -6a & 0\\
C_+ & \ep_0-\be - 2 a & 2 C_+ & -2a \\
0 & 2 C_+ & \ep_0-2 \be -2a& C_+\\
0&0&3C_+&\ep_0-3\be \end{pmatrix}
\end{displaymath}
with
$$
\be = 2(a+ B) +1\,.
$$
The eigenfunctions of $\Hb_3|_{\M_1}$ are
$$
\chi_j(\t) = k_{j0} + k_{j1}\,\t_1 + k_{j2}\,\t_2 + \t_3\,,\qquad
0\le j\le 3\,,
$$
where
\begin{align*}
k_{j0}&=\frac{1}{2 C_+ \E_j} \left[(\E_j - 2a + 2 \be)(\E_j+ 3
\be)-3 C_+^2\right]\,,\\ k_{j1}&=\frac1{6 C_+^2}\left[(\E_j + 2a + 2
\be)(\E_j+ 3 \be)-3 C_+^2 \right]\,,\\ k_{j2} &= \frac{\E_j+ 3 \be}{3 C_+}\,,
\end{align*}
and $\E_j$ is one of the four roots of the characteristic polynomial
\begin{multline*}
\E^4 + 2(2a + 3\be) \E^3+ (4 a^2+ 18 a \be + 11
\be^2 - 10 C_+^2) \E^2 \\ {}+
6(\be^3 + 3 a \be^2 + 2 a^2 \be+2 a C_+^2- 5\be C_+^2) \E
+ 9 C_+^2(-2\be^2+2a\be+C_+^2)\,.
\end{multline*}
The corresponding eigenfunctions $\psi_j(\x)=\mu(\bz)\,\chi_j(\t)$
($0\le j\le 3$) of $H_3$ have energy $E_j=\E_j+\ep_0$. If the parameters
$a$, $B$ and $C_+$ take the values
$$
a=2\,,\quad B=\frac32\,,\quad C_+ = -1 \,,
$$
compatible with the normalizability of the algebraic eigenfunctions,
then the eigenvalues can be calculated numerically:
\begin{align*}
E_0 &= -69.7926 \\ E_1&= -64.1121\\ E_2&= -56.4954 \\E_3&=
-44.5999\,.
\end{align*}
The gauge part of the algebraic eigenfunctions is
given by
\begin{align*}
    \chi_0(\t)&= 0.211595 +0.376202\,\t_1 + 
    0.347522\,\t_2 + \t_3\\
    \chi_1(\t)&=-0.959207 - 0.00694338\,\t_1 - 
    1.54596\,\t_2 + \t_3 \\
    \chi_2(\t)&=0.00509379 + 16.3594\,\t_1 - 
    4.08486\,\t_2 + \t_3\\
    \chi_3(\t)&= -967.257  + 80.6047\,\t_1 - 
    8.05004\,\t_2 + \t_3\,.
\end{align*}
Since the symmetric variables
\begin{align*}
 \t_1&=\cosh x_1+\cosh x_2+\cosh x_3\\
 \t_2&=\cosh x_1\,\cosh x_2+\cosh x_1\,\cosh x_3+\cosh x_2\,\cosh
 x_3\\
 \t_3&=\cosh x_1\,\cosh x_2\,\cosh x_3
\end{align*}
are all positive, $\chi_0$, and therefore $\psi_0$, does not vanish in
the region \eqref{region2}. This shows that $\psi_0$ is the ground
state of the potential \eqref{QESex}.

In the QES cases we are able to compute a few energies and their
corresponding eigenfunctions algebraically, but we know nothing about
the rest of the spectrum. In fact, it is natural to expect that the
potential of this example has some levels between the ones we have
obtained.
\subsection{ES potential from Case 3}
For the ES potential \eqref{vn3es}
\begin{equation}
    V_N(\x)=\sum_k\left[\frac{C_-^2}4\,\e^{-2x_k}+
    \frac12 C_0\,C_-\e^{-x_k}+
    \right] +\frac{g_2}4\sum_{j\ne
    k}\csch^2\left(\frac{x_j-x_k}2\right)\,,
    \label{vn3ESex}
\end{equation}
the gauge Hamiltonian in the symmetric variables is given by
\begin{align}
    \Hb_N = &-\sum_{i,j}\left[j\,\t_i\t_j+\sum_{k=1}^j
    (j-i-2k)\t_{i+k}\t_{j-k}\right]\d_{\t_i}\d_{\t_j}\notag\\
    &{}-\sum_j j\left[2B+1+a(2N-j-1)\right]\t_j\,\d_{\t_j}\notag\\
    &{}-C_-\sum_j(N-j+1)\t_{j-1}\d_{\t_j}
    +E_0\,,
    \label{hbnex3}
\end{align}
with
\begin{equation}
B=\frac12(C_0+1)
\label{Bdefex}
\end{equation}
and $E_0$ given by \eqref{gstate}.
The gauge Hamiltonian \eqref{hbnex3} preserves each subspace
$\hat{\mathcal M}_k$ for arbitrary $k$. For instance, the
restriction of $\Hb_N$ to $\hat{\mathcal M}_2$ has the following
matrix
with respect to the basis $\left\{1,\t_1,\t_2,\t_1^2\right\}$ of
$\hat{\mathcal M}_2$:
\begin{equation}
    \begin{pmatrix}
    E_0 & -NC_- & 0 & 0\\
    0 & E_0+1-\be & -(N-1)C_- & -2NC_-\\
    0 & 0 & E_0+2(a-\be+1) & 4\\
    0 & 0 & 0 & E_0-2\be
    \end{pmatrix},
    \label{hb3mat}
\end{equation}
where $\be$ is defined by Eqs.~\eqref{betaex} and \eqref{Bdefex}. By
diagonalizing this matrix, we can again exactly compute four
eigenfunctions with their corresponding energies for the multiparticle
potential \eqref{vn3ESex}. The energies can be simply read off the
diagonal elements of the matrix \eqref{hb3mat}, and coincide with
those of the previous example (with $B$ now given by \eqref{Bdefex}).
The corresponding eigenfunctions are given by
\begin{equation}
    \psi_i=\mu\,\chi_i\,,\qquad 0\le i\le 3\,,
    \label{eigex3}
\end{equation}
where
\begin{equation}
    \mu=\prod_{j<k}\sinh^{\,a}\left(\frac{x_j-x_k}2\right)\,
    \cdot
    \exp\left\{\sum_k\left[\frac12\bigl(C_0+1 + a(N-1)\bigr)\,x_k
    -\frac{C_-}2\e^{-x_k}\right]\right\},
    \label{muex3}
\end{equation}
and
\begin{align}
    \chi_0&=1\,,\\
    \chi_1&=\t_1+\frac{N C_-}{\be-1}\,,\\
    \chi_2&=\t_1^2-\frac{2\,\t_2}{a+1}
    +\frac{2C_-(Na+1)}{(a+1)(\be+1)}\,\t_1
    +\frac{N(aN+1)C_-^2}{\be(a+1)(\be+1)}\,,\\
    \chi_3&=\t_2+\frac{C_-(1-N)}{2a-\be+1}\,\t_1
    +\frac{N(N-1)C_-^2}{2(a-\be+1)(2a-\be+1)}\,,
\end{align}
with
\begin{equation}
    \t_1=\sum_k e^{x_k}\,,\qquad
    \t_2=\sum_{j<k}e^{x_j+x_k}\,.
    \label{t12ex3}
\end{equation}
Again, $\psi_0=\mu$ has no zeros in the configuration space
\eqref{region1}, and is therefore the ground state of the system. As
in the previous example, we have exactly computed some eigenvalues and
eigenfunctions of the potential \eqref{vn3ESex} for an arbitrary
number of particles.
\subsection{QES potential from Case 3}
As a final example, consider the QES potential \eqref{vn3q1} with 
$m=1$
\begin{align}
    V_N(\x)=&\sum_k\left[\frac{C_-^2}4\,\e^{-2x_k}+
    \frac12 C_0\,C_-\e^{-x_k}+
    C_+\bigl(a(N-1)+\frac{C_0}2+2\bigr)\e^{x_k}+
    \frac{C_+^2}4\,\e^{2x_k}\right]
    \notag\\
    &\quad{}+\frac{g_2}4\sum_{j\ne k}\csch^2\left(\frac{x_j-x_k}2\right).
    \label{vn3qm1}
\end{align}
The restriction of the gauge Hamiltonian $\Hb_N$ to
the subspace
$$
\M_1=\text{span}\left\{1,\t_1,\dots,\t_N\right\}
$$
is given in terms of the symmetric variables $\t_j$ by the following 
expression
\begin{multline}
    \Hb_N|_{\M_1}=
    \ep_0+C_+\,\t_1\left(1-\sum_j \t_j\,\d_{\t_j}\right)+
    C_+\sum_j(j+1)\,\t_{j+1}\d_{\t_j}\\
    -\sum_j\left[C_0+a(2N-j-1)+2\right]j\,\t_j\,\d_{\t_j}
    -C_-\sum_j(N-j+1)\,\t_{j-1}\,\d_{\t_j}\,,
    \label{hg31}
\end{multline}
with
\begin{equation}
    \ep_0=-\frac N4\left[\bigl(C_0+a\,(N-1)+1\bigr)^2+\frac{a^2}{3}\,(N^2-1)
    +2\,C_-C_+\right].
    \label{ep0}
\end{equation}
The matrix $\left(h_{ij}\right)_{0\le i,j\le N}$ of $\Hb_N|_{\M_1}$
with respect to the basis $\left\{1\equiv\t_0,\t_1,\dots,\t_N\right\}$
of $\M_1$ is therefore tridiagonal, with nontrivial elements
\begin{align}
    &h_{j,j-1}=j\,C_+\,,\qquad
    h_{j,j+1}=-C_-(N-j)\,,
    \label{hextrad}\\
    &h_{jj}=\ep_0-j\left[C_0+a(2N-j-1)+2\right]\equiv h_j\,.
    \label{hdiag}
\end{align}

The tridiagonal character of the matrix $\left(h_{ij}\right)$ has
important consequences for the calculation of the eigenfunctions of
$\Hb_N$ lying in $\M_1$. Indeed, if we write one such eigenfunction
with eigenvalue $E$ as
\begin{equation}
    \chi_E(\t) = \sum_{j=0}^N \tp_j(E)\,\t_j\,,
    \label{chiE}
\end{equation}
it follows from Eqs.~\eqref{hextrad}--\eqref{hdiag} that the 
coefficients $\tp_j(E)$ satisfy the following
relations
\begin{equation}
    j C_+\,\tp_{j-1}(E)+(h_j-E)\tp_j(E)-C_-(N-j)\tp_{j+1}(E)=0\,,
    \qquad j=0,1,2,\dots,N\,,
    \label{rr}
\end{equation}
where we can take $\tp_{-1}=0$ and, without loss of generality,
$\tp_0=1$. Let us now regard the energy $E$ in Eq.~\eqref{rr} as a
real variable, and let us consider the recurrence relation
\eqref{rr} for arbitrary values of $j\in\NN$. Following the usual
procedure, we introduce new normalized coefficients $\p_j(E)$
through the formula
\begin{equation}
    \p_j(E) = \frac{N!}{(N-j)!}\,(-C_-)^j\,\tp_j(E)\,.
    \label{pij}
\end{equation}
Note that $\tp_0(E)=1$ implies that $\p_0(E)=1$. The coefficients
$\p_j(E)$ then satisfy the following three-term recurrence relation in
canonical form
\begin{equation}
    \p_{j+1}(E)=(E-h_j)\,\p_j(E)-\a_j\,\p_{j-1}(E)\,,
    \label{rrcan}
\end{equation}
with
\begin{equation}
    \a_j = -j(N-j+1)C_-C_+\,.
    \label{ajdef}
\end{equation}
The functions $\p_j(E)$ defined by the
latter relation with the initial conditions $\p_{-1}(E)=0$,
$\p_0(E)=1$ are monic polynomials of degree $j$. It is well known that
the canonical form of the recurrence relation \eqref{ajdef} entails
that the polynomials $\p_j(E)$ are an orthogonal family with respect
to a suitably defined weight functional, \cite{Ch78}. Furthermore, the
fact that the coefficient $\a_j$ vanishes for $j=N+1$ implies that
this functional is not positive definite; in fact, the polynomials
$\pi_j$ with $j\ge N+1$ must have zero norm. Therefore, the polynomial
family $\left\{\p_j(E)\right\}_{j\ge0}$ is \emph{weakly} orthogonal,
cf.~\cite{Ch78}. We have thus associated in a natural way a weakly
orthogonal polynomial family to the QES many-body potential
\eqref{vn3qm1}. Recall, in this respect, that it is possible to 
construct a weakly orthogonal polynomial family for (almost) every QES 
one-particle potential on the line, \cite{FGRorth96}.

Going back to the calculation of the eigenfunctions of $\Hb_N$
belonging to $\M_1$, note first of all that their energies are the
zeros of the \emph{critical polynomial}
$\p_{N+1}(E)$. This can be seen, for instance, by observing that 
the $j$-th principal minor $\de_j(E)$ of the matrix $E-(h_{ij})$ 
satisfies the same recurrence relation as $\pi_j(E)$, with the same 
initial conditions, so that
\begin{equation}
    \de_j(E)=\p_j(E)\,,
    \label{minors}
\end{equation}
and, in particular,
\begin{equation}
    \de_{N+1}(E)=\det\bigl(E-(h_{ij})\bigr)=\p_{N+1}(E)\,.
    \label{charp}
\end{equation}
Since $C_+C_-<0$ for the potential \eqref{vn3qm1}, it follows from
\eqref{ajdef} that $\a_j$ is positive for $j=1,2,\dots,N$. This
implies, \cite{Ar64}, that the critical polynomial $\p_{N+1}(E)$ has
$N+1$ distinct real roots $E_0<E_1<\dots<E_N$. By the previous
equation, the spectrum of the restriction of $\Hb_N$ to $\M_1$
consists of the $N+1$ real eigenvalues $E_0<\dots<E_N$. The
respective (unnormalized) eigenfunctions $\chi_i\equiv\chi_{E_i}$ are
given by
\begin{equation}
    \chi_i(\t)=1+\frac1{N!}\,\sum_{j=1}^N 
    (-1)^j\,\frac{(N-j)!}{C_-^j}\,\pi_j(E_i)\,\t_j\,,
    \label{eig3qes}
\end{equation}
where the coefficients $\pi_j(E_i)$ are calculated either from the
recurrence relation \eqref{rrcan} (for $E=E_i$), or by computing the
minors $\de_j(E_i)$. The corresponding eigenfunctions of the physical
Hamiltonian $H_N$ are obtained by multiplying each function $\chi_i$
in \eqref{eig3qes} by the gauge factor
\begin{equation}
    \mu=\prod_{j<k}\sinh^{\,a}\left(\frac{x_j-x_k}2
    \right)\,
    \cdot 
    \exp\left\{\frac12\sum_k\left[\bigl(C_0 + a(N-1)+1\bigr)x_k+
    C_+\e^{x_k}-C_-\e^{-x_k}\right]
    \right\}.
    \label{g3qes}
\end{equation}

We can also express the eigenvalues of $\Hb_N|_{\M_1}$ and 
their eigenfunctions using continued fractions, by working with the 
quotients
\begin{equation}
    q_j(E)=\frac{\pi_j(E)}{\pi_{j-1}(E)}\,,\qquad j=1,2,\dots\,.
    \label{qjdef}
\end{equation}
Then Eq.~\eqref{rrcan} becomes
\begin{equation}
    q_{j+1}(E) = E-h_j-\frac{\a_j}{q_j(E)}\,,
    \label{qrr}
\end{equation}
from which we immediately obtain the continued fraction expansion
\begin{equation}
    q_{j+1}(E)=E-h_j-\frac{\a_j}{E-h_{j-1}-}\;\frac{\a_{j-1}}{E-h_{j-2}-}
    \;\dots\;\frac{\a_2}{E-h_1-}\;\frac{\a_1}{E-h_0}\,.
    \label{cfe}
\end{equation}
The eigenvalue equation in this formalism can be obtained by imposing the
vanishing of $q_{N+1}(E)$ (since it is proportional to $\pi_{N+1}(E)$, 
by Eq.~\eqref{qjdef}),
that is
\begin{equation}
    E-h_N=\frac{\a_N}{E-h_{N-1}-{}}\;\frac{\a_{N-1}}{E-h_{N-2}-{}}
    \;\dots\;\frac{\a_2}{E-h_1-{}}\;\frac{\a_1}{E-h_0}\,.
    \label{eigcfn}
\end{equation}
Alternatively, solving Eq.~\eqref{qrr} for $q_j$ we obtain
\begin{equation}
    q_j(E) = \frac{\a_j}{E-h_j-q_{j+1}(E)}\,,
    \label{qjj+1}
\end{equation}
and an equivalent form of the eigenvalue equation \eqref{eigcfn}
follows by expressing $q_1(E)=E-h_0$ as a continued fraction:
\begin{equation}
    E-h_0 = \frac{\a_1}{E-h_1-{}}\;\frac{\a_2}{E-h_2-{}}
    \;\dots\;\frac{\a_{N-1}}{E-h_{N-1}-{}}\;\frac{\a_N}{E-h_N}
    \label{eigcf0}
\end{equation}
(since $q_{N+1}(E)$ vanishes if and only if $E$ is an eigenvalue). The eigenfunctions are still given by \eqref{eig3qes}, where of 
course now
\begin{equation}
    \pi_j(E)=\prod_{i=1}^j q_i(E)\,.
    \label{pijqj}
\end{equation}

Consider, as an example, the potential \eqref{vn3qm1} with $N=3$ 
particles, for which
\begin{equation}
    \ep_0=-\frac34(C_0+2a+1)^2-2a^2-\frac32C_-C_+\,.
    \label{ep0exa3}
\end{equation}
The eigenfunctions of $\Hb_3|_{\M_1}$ can be expressed as
\begin{equation}
    \chi_j(\t)=1+k_{j1}\,\t_1+k_{j2}\,\t_2+k_{j3}\,\t_3\,,
    \label{chijex3}
\end{equation}
where the coefficients $k_{jl}=\tp_l(E_j)$ are given by
\begin{align}
k_{j1} &= -\frac{\E_j}{3C_-}\,,\\
k_{j2} &= \frac1{6C_-^2}\left[\E_j^2+(2a+\be)\E_j+3C_-C_+\right],\\
k_{j3} &= -\frac1{6C_-^3}
\left[\E_j^3+(4a+3\be)\E_j^2+(4a^2+6a\be+2\be^2+7C_-C_+)\E_j
+6(a+\be)C_-C_+\right].
\end{align}
In the latter equations we have set
\begin{equation}
    \be= 2a+C_0+2\,,
    \label{}
\end{equation}
and $\E_j=E_j-\ep_0$ is one of the roots of the polynomial
\begin{multline}
    \p_4(\E+\ep_0)=\E^4+(4a+6\be)\E^3+(4a^2+18a\be+11\be^2+10C_-C_+)\,\E^2\\
    {}+6\bigl(\be(2a^2+3a\be+\be^2)+(2a+5\be)C_-C_+\bigr)\,\E
    +9C_-C_+\bigl(2\be(a+\be)+C_-C_+\bigr)\,.
    \label{chpexa3}
\end{multline}
For instance, if the parameters in the potential are given by
\begin{equation}
    a=2\,,\qquad C_-=2\,,\qquad C_0=1\,,\qquad C_+=-1
    \label{param3qes}
\end{equation}
then
$$
\ep_0=-32
$$
and
\begin{align*}
    E_0 &= -54.5584\\
    E_1 &= -49.5497\\
    E_2 &= -42.4323\\
    E_3 &= -31.4596\,.
\end{align*}
The eigenfunctions of $\Hb_3$ in $\M_1$ are
\begin{align*}
    \chi_0(\t) &= 1+3.75974\,\t_1 + 10.6142\t_2 + 20.4323\,\t_3\\
    \chi_1(\t) &= 1+2.92495\,\t_1 + 4.5394\,\t_2 - 3.94697\,\t_3\\
    \chi_2(\t) &= 1+1.73871\,\t_1 - 0.496778\,\t_2 + 0.141027\,\t_3\\
    \chi_3(\t) &= 1-0.090071\,\t_1 + 
    0.00986437\,\t_2 - 0.00137384\,\t_3\,.
\end{align*}
Since the symmetric variables, given in this case by \eqref{t12ex3} and
\begin{equation}
    \t_3 = \sum_{i<j<k}\e^{x_i+x_j+x_k}\,,
    \label{sym3}
\end{equation}
are all positive, the function $\psi_0=\mu\chi_0$ is once again
the ground state of the Hamiltonian $H_3$.
%
%
\section{Summary and Conclusions}
\label{sumcon}
We have used Calogero's construction of classical solvable many-body
systems and applied it to the most general one-dimensional
quasi-exactly solvable normalizable Schr\"odinger operator in the
line. The corresponding quantum many-body Hamiltonians have been shown
to have an algebraic structure that enables us to compute part or all
of their spectrum by straightforward algebraic means. In all cases, if
the one-dimensional \emph{seed potential} is ES (QES) then the
corresponding many-body potential constructed on it is also ES (QES).

The QES examples include a generalization of the sextic deformation of
Olshanetsky--Perelomov's $B_N$ rational model recently found in
\cite{Shifman}, as well as some new deformations of the standard
hyperbolic $BC_N$ potential. We have also found an additonal QES
deformation of the $B_N$ rational model not related to the one
considered in Ref.~\cite{Shifman}.

The ES cases include $A_N$ and $B_N$ rational models with harmonic
force and some hyperbolic OP potentials of $BC_N$ type, as
well as the $A_N$ hyperbolic model with an external Morse potential
discussed by Inozemtsev and Meshcheryakov, \cite{Inocencio}. Since we
have only looked for Hamiltonians having normalizable eigenfunctions,
the models treated in \cite{OP} with no discrete spectrum do not appear
in this work. The fact that we have considered only non-periodic
one-dimensional seed potentials, \cite{Artemio94}, explains why
the trigonometric integrable Hamiltonians such as Sutherland's model
are also absent. In a future work we plan to perform a similar
analysis using \emph{periodic} seed potentials which could lead to new
results on the elliptic case, where much less is known.

The ES problems have the additional feature that their
associated gauge Hamiltonian $\Hb_N$ preserves not only the
subspaces $\mathcal M_k$ but the smaller subspaces $\hat \mathcal
M_k$ defined by \eqref{Lk}. This implies that the number of particles $N$
represents no difficulty in these cases, and allows us to
calculate some levels and eigenfunctions for arbitrary $N$.
Moreover, we have shown that an ordered basis in each subspace
$\hat \mathcal M_k$ can be chosen so that the action of $\Hb_N$ is
triangular. As a consequence, the diagonal terms automatically
give the discrete spectrum, and recursive expressions for the
eigenfunctions can in principle be written.

It should be stressed that the results we have obtained depend greatly
on the ansatz \eqref{gfactor} for the gauge factor $\mu$. Our ansatz
has been chosen by analogy with the one-dimensional case, but it is
clear that more general ans\"atze could lead to more general results.
In his recent work \cite{Cal3}, Calogero investigates a very general
ansatz which gives rise to higher-body interactions. A systematic
analysis of this method should focus on the equivalence problem,
attempting to solve it in full generality.
%
%
\section{Acknowledgments}
\label{ack}
It is our pleasure to thank F. Calogero, A. Perelomov, and O. Ragnisco
for useful discussions. D. G.-U. is also glad to acknowledge F.
Calogero's warm hospitality at Universit\'a di Roma (La Sapienza)
while this work was in progress.
\section{Appendix}
\label{app}
The following expressions are useful in the transition from the
canonical coordinates $\bf z$ to the symmetric variables $\t$.
(All indices  run from $1$ to $N$, with the restrictions
indicated under the summation symbol.)
\begin{displaymath}
 \Lambda_1^{(p)} \equiv \sum_{k\neq j}
\frac{z_k^p}{z_k-z_j}
\end{displaymath}
\begin{align}
 \Lambda_1^{(0)}&= 0 \label{L10}\\[3pt]
 \Lambda_1^{(1)}&= \frac{1}{2}N(N-1) \label{L11}\\[3pt]
\Lambda_1^{(2)}&= (N-1) \t_1 \label{L12}
\end{align}
\vskip .5cm
\begin{displaymath}
\Lambda_2^{(p)} \equiv \sum_{j\neq k \neq l\ne j}
\frac{z_k^p}{(z_k-z_j) (z_k-z_l)}
\end{displaymath}
\begin{align}
\Lambda_2^{(0)}&= \Lambda_2^{(1)}= 0 \label{L20}\\[3pt]
\Lambda_2^{(2)}&= \frac{1}{3} N(N-1)(N-2) \label{L22}
\end{align}
\vskip .5cm
The following expressions are useful to express
differential operators in terms of the symmetric variables (it is 
understood that $\t_k=0$ for $k>N+1$):
\begin{displaymath}
\sum_j z_j^p\,\d_{z_j} = \sum_j B_j^{(p)}\,\d_{\t_j}
\end{displaymath}
\begin{align}
B_j^{(0)} &= (N-j+1)\t_{j-1} \label{Bj0}\\[2pt] B_j^{(1)} &= j \t_j
\label{Bj1}\\[2pt]B_j^{(2)} &= \t_1 \t_j - (j+1) \t_{j+1}
\label{Bj2}
\end{align}
\vskip .5cm
\goodbreak
\begin{displaymath}
\sum_k z_k^p\,\d^2_{z_k} = \sum_{i,j} A_{ij}^{(p)}\,\d_{\t_i}
\d_{\t_j}
\end{displaymath}
\begin{align}
A_{ij}^{(0)} &= (N-i+1)\t_{i-1} \t_{j-1}-
\sum_{k=1}^{j-1}(i-j+2k)\t_{i+k-1}\t_{j-k-1} \label{Aij0}
\\[2pt] A_{ij}^{(1)} &=
\sum_{k=0}^{j-1}(i-j+2k+1)\t_{i+k}\t_{j-k-1} \label{Aij1}
\\[2pt] A_{ij}^{(2)} &= j \t_i \t_j + \sum_{k=1}^{j} (j-i-2k)\t_{i+k}\t_{j-k}
\label{Aij2}
\end{align}
\vskip .5cm
\begin{displaymath}
2\sum_{j\neq k} \frac{z_k^p}{z_k-z_j}\,\d_{z_k} = \sum_j
C_j^{(p)}\,\d_{\t_j}
\end{displaymath}
\begin{align}
C_j^{(0)} &= -(N-j+1)(N-j+2)\t_{j-2}\label{Cj0}\\[2pt]
C_j^{(1)} &= (N-j)(N-j+1) \t_{j-1}\label{Cj1}\\[2pt]
C_j^{(2)} &= j(2N-j-1) \t_j \label{Cj2}
\end{align}
%
%
\goodbreak


\begin{thebibliography}{99}
\frenchspacing
\bibitem {Cal1} Calogero F 1969 \emph{J. Math. Phys.} {\bf 10}
2191--7
\bibitem{Cal2} Calogero F 1971 \emph{J. Math. Phys.} {\bf 12} 419--36
\bibitem{Suth} Sutherland B 1971 \emph{Phys. Rev.} {\bf A4}
2019--21
\bibitem{OPclass} Olshanetsky M A and Perelomov A M 1981 \emph{Phys.
Rep.} {\bf 71} 313--400
\bibitem{OP} Olshanetsky M A and Perelomov A M 1983 \emph{Phys.
Rep.} {\bf 94} 313--404
\bibitem{Turb1} R\"uhl W and Turbiner A 1995 \emph{Mod. Phys. Lett.}
{\bf 10} 2213--21
\bibitem{Turb3} Brink L, Turbiner A and Wyllard N 1998 \emph{J.
Math. Phys} {\bf 39} 1285--315
\bibitem{GKOJPA} Gonz\'alez-L\'opez A, Kamran N and Olver P J 1991
\emph{J. Phys. A: Math. Gen.} {\bf 24} 3995--4008
\bibitem{Caltruco} Calogero F 1978 \emph{Nuovo Cim.} {\bf 43B}
177--241
\bibitem{Shifman} Hou X and Shifman M 1999 \emph{Int.~J. Mod.~Phys.}
{\bf A14} 2993--3004
\bibitem{Inocencio} Inozemtsev V I and Meshcheryakov D V 1986
\emph{Physica Scripta} {\bf 33} 99--104
\bibitem{Artemio93}  Gonz\'alez-L\'opez A, Kamran N and Olver P J 1993
\emph{Commun. Math. Phys.} {\bf 153} 117--46
\bibitem{Inocencio2} Inozemtsev V I 1984
\emph{Physica Scripta} {\bf 29} 518--20
\bibitem{InoMesLMP} Inozemtsev V I and Meshcheryakov D V 1985
\emph{Lett. Math. Phys.} {\bf 9} 13--8
\bibitem{Inocencio3} Inozemtsev V I and Meshcheryakov D V 1984
\emph{Phys. Lett.} {\bf 106A} 105--8
\bibitem{Turb2} Minzoni A, Rosenbaum M and Turbiner A 1996
\emph{Mod. Phys. Lett.} {\bf A24} 1977--84
\bibitem{Ch78} Chihara T S 1978 {\em An Introduction to Orthogonal
Polynomials} (New York: Gordon and Breach)
\bibitem{FGRorth96}
Finkel F, Gonz\'alez-L\'opez A, and Rodr\'\i guez M A 1996 {\em J. Math.
Phys.} {\bf 37} 3954--72
\bibitem{Ar64}
Arscott F M 1964 {\em Periodic Differential Equations} 
(Oxford: Pergamon)
\bibitem{Artemio94} Gonz\'alez-L\'opez A, Kamran N and Olver P J 1994
\emph{Contemporary Mathematics} {\bf 160} 113--40
\bibitem{Cal3} Calogero F 1999 \emph{J. Math. Phys.} {\bf 40}
4208--26
\end{thebibliography}
\end{document}